\documentclass[useAMS,usenatbib]{mn2e}
\usepackage{graphics}
\usepackage{graphicx}
\input{epsf}
\begin{document}

\def\bc{\begin{center}}
\def\ec{\end{center}}
\def\b{\begin{equation}}
\def\e{\end{equation}}
\def\ber{\begin{eqnarray}}
\def\eer{\end{eqnarray}}
\def\l{\left}
\def\r{\right}
\def\eg{{\it e.g.}}
\def \ie {{\em i.e.~~}}
\def \lleq {\lower0.9ex\hbox{ $\buildrel < \over \sim$} ~}
\def \ggeq {\lower0.9ex\hbox{ $\buildrel > \over \sim$} ~}
\def \dlt {\delta}
\def \L {\Lambda}
\def \T {\tau}
\def \l {\Lambda}

\title[Measuring the Geometry and Topology of Large Scale Structure using 
        SURFGEN]{Measuring the Geometry and Topology of Large Scale Structure
  using SURFGEN: Methodology and Preliminary Results}
\author[J.V. Sheth, V. Sahni, S.F. Shandarin, B.S.
Sathyaprakash]{Jatush V.Sheth$^{1,4}$, Varun Sahni$^{1,5}$,
  Sergei F.Shandarin$^{2,6}$ and B.S.Sathyaprakash$^{3,7}$ \\
  $^{1}$ Inter University Centre for Astronomy $\&$ Astrophysics, Pune, India \\
  $^{2}$ Department of Physics and Astronomy, University of Kansas,KS
  66045, USA \\
  $^{3}$Department of Physics and Astronomy, University of Cardiff,
  Cardiff, UK \\
  $^{4}$jvs@iucaa.ernet.in \\
  $^{5}$varun@iucaa.ernet.in \\
  $^{6}$ sergei@ukans.edu,sergei@ku.edu \\
  $^{7}$B.Sathyaprakash@astro.cf.ac.uk \\
  } \maketitle
%%%%%%%%%%%%%%%%%%%%%%%%%%%%%%%%%%%%%%%%%%%%%%%%%%%%%%%%%%%%%%%%%%%%%%%%%%%%%%
\begin{abstract}
  Observations of the universe reveal that matter within it clusters
  on a variety of scales. On scales between 10 - 100 Mpc, the universe
  is spanned by a percolating network of superclusters interspersed
  with large and almost empty regions -- voids.  This paper, the first
  in a series, presents a new ansatz which can successfully be used to
  determine the morphological properties of the supercluster-void
  network. The ansatz is based on a surface modelling scheme
  (SURFGEN), developed explicitly for the purpose, which generates a
  triangulated surface from a discrete data set representing (say) the
  distribution of galaxies in real (or redshift) space.  The
  triangulated surface describes, at progressively lower density
  thresholds, clusters of galaxies, superclusters of galaxies and
  voids.  Four {\em Minkowski functionals} (MFs) -- surface area,
  volume, extrinsic curvature and genus -- describe the geometry and
  topology of the supercluster-void network.  On a discretised and
  closed triangulated surface the MFs are determined using SURFGEN.
  Ratio's of the Minkowski functionals provide us with an excellent
  diagnostic of three dimensional shapes of clusters, superclusters
  and voids. Minkowski functionals can be studied at different levels
  of the density contrast and therefore probe the morphology of large
  scale structure on a variety of length scales.  Our method for
  determining the Minkowski functionals of a triangulated iso-density
  surface is tested against both simply and multiply connected eikonal
  surfaces such as triaxial ellipsoids and tori.  The performance of
  our code is thereby evaluated using density distributions which are
  pancake-like, filamentary, ribbon-like and spherical.  Remarkably,
  the first three Minkowski functionals are computed to {\em better
    than 1$\%$} accuracy while the fourth (genus) is known exactly.
  SURFGEN also gives very accurate results when applied to Gaussian
  random fields.  We apply SURFGEN to study morphology in three
  cosmological models, $\l$CDM, $\T$CDM and SCDM, at the present
  epoch. Geometrical properties of the supercluster-void network are
  found to be sensitive to the underlying cosmological parameter set.
  For instance, the percolating supercluster in $\l$CDM turns out to
  be more filamentary but topologically simpler than superclusters in
  $\T$CDM and SCDM.  It occupies just $0.6\%$ of the total
  simulation-box volume yet contains about $4\%$ of the total mass.
  Our results indicate that the surface modelling scheme to calculate
  Minkowski functionals is accurate and robust and can successfully be
  used to quantify the topology and morphology of the
  supercluster-void network in the universe.
\end{abstract}

\begin{keywords}
  cosmology: theory---structure
  formation---morphology---superclusters---voids--cosmic web
\end{keywords}

\section{Introduction}
The existence of the supercluster-void network is one of the most
intriguing observational features of our universe.  Redshift surveys
of galaxies confirm that on very large scales the bulk of matter in
the universe is concentrated in clusters and superclusters of galaxies
which are separated by large almost empty regions, appropriately
called voids.  At moderate density thresholds $\delta \sim 1$, the
supercluster network percolates even though it occupies a tiny
fraction of the total volume, and so has a small filling fraction.

The morphology of the supercluster-void network is quite complex and
has inspired evocative descriptions such as being `honey-comb-like',
`bubble-like', `a filamentary network', `swiss cheese', `cosmic web'
etc.  Indeed the intricate weave of large scale structure is becoming
increasingly evident as results from progressively larger galaxy
redshift surveys (CfA, IRAS, PSCz, LCRS, 2dFGRS, SDSS) demonstrate.
Quantifying the properties of the supercluster-void network is clearly
one of the cardinal tasks facing cosmology today.  Since the network
evolved from an almost featureless Gaussian random field (as suggested
by observations of the cosmic microwave background) it is important to
understand how such a rich and complex cosmic tapestry --
characterized by prominent non-Gaussian features (clusters,
superclusters, voids) -- could have arisen from small and
statistically random initial conditions.  Furthermore, since the
network is highly evolved it is unlikely that its principle features
can be described within a strictly perturbative framework which breaks
down when $\delta \ggeq 1$. It is therefore both interesting and
revealing that approximations which successfully describe non-linear
dynamics (such as the Zeldovich approximation and the Adhesion model)
generically predict the formation of filaments and pancakes, which
interweave and percolate to form the large scale structure of the
universe % \citep{shandzed89}, \citep{gurbatov85}, \citep{gurbatov89}.
\citep{shandzed89, gurbatov85, gurbatov89}.  Furthermore, this
interweaving network of filaments, sheets and voids -- predicted both
by the Zeldovich approximation and the adhesion model -- is readily
seen in N-body simulations of gravitational clustering for a wide
variety of initial power spectra \citep{ksh93,sss96}. It therefore
appears that a filamentary distribution of large scale structure is an
almost generic outcome of pressureless (dark) matter clustering from
Gaussian initial conditions.  (A theoretical model for understanding
the morphology of large scale structure can be found in the ``cosmic
web'' hypothesis of \citet{bondo}.  Insightful overviews of the
statistical and dynamical techniques used in studies of large scale
structure can be found in: \citet{sc95,masaar,rien}.)

A number of statistical measures have been advanced to quantify the
pattern made by galaxies as they cluster in our universe.  Prominent
among them are Percolation Analysis \citep{zes82,shz83}, Counts in
Cells \citep{jd83, lapp91}, Minimal Spanning Trees \citep{barrow85},
the Genus measure \citep{gott86} etc.  (To some extent these methods
complement the traditional approach to quantify clustering using the
hierarchy of correlation functions \citep{peeblesbook80}, which become
cumbersome to evaluate beyond the three point function.)  Recently
Mecke, Buchert and Wagner (1994) have introduced the Minkowski
functionals (hereafter MFs) to cosmology. MFs contain information
pertaining to geometry, connectivity (percolation) and topology
(genus). In addition, the ratio's of Minkowski functionals quantify
the {\em morphology} of large scale structure.  This has led to the
construction of a set of measures called Shapefinders which tell us
whether the distribution of matter in superclusters/voids is
spherical, planar, filamentary etc. \citep{sss98,sss98b}.
(Applications of Shapefinders to galaxy catalogues and N-body
simulations has been discussed in \citet{basilakos,kolokotronis}.
Moment-based studies of morphology predating the Shapefinders can be
found in \citet{bs92,lv95}.)  Thus between them, the four Minkowski
functionals contain valuable information regarding both the
geometrical as well as topological distribution of matter in the
universe. Since Minkowski functionals are additive in nature one can
glean information regarding both individual objects (galaxies,
clusters, voids) as well as describe the supercluster-void network in
its totality.

There have been three major attempts made to study the morphology of
the LSS using MFs; these efforts differ in their approach of
evaluating the MFs: (1) Firstly, Boolean grain models study the MFs of
surfaces which result due to intersecting spheres decorating the input
point-set (Mecke, Buchert $\&$ Wagner 1994). (2) Secondly, Krofton's
formulae make it possible to calculate MFs on a density field defined
on a grid (Schmalzing $\&$ Buchert 1997). In this case the MFs are
calculated by using the information of the number of cells in 1D
(vertices), 2D (faces) and 3D (cuboids). (3) Finally, an alternative,
resolution-dependent approach consists in employing the Koenderink
invariants (Schmalzing $\&$ Buchert 1997, Schmalzing et al. 1999).

This paper introduces a radically new and powerful approach for
computing the Minkowski functionals. This approach consists of
constructing iso-density surfaces using an elaborate surface modelling
scheme defined in terms of excursion sets of a density field. These
surfaces are triangulated and the MFs are evaluated for the resulting
closed polyhedral surface.  A similar algorithm has been used in
studies of two-dimensional CMB maps \citep{nfs99,shetal02}. However,
developing a three-dimensional algorithm requires a new component that
builds a surface of approximately constant (to linear order) density.
It has been implemented and reported in this paper. 

The MFs so constructed are tested against known results for simply
connected surfaces (triaxial ellipsoids) as well as multiply connected
surfaces (triaxial tori).  In all cases we find that our ansatz for
determining MFs on polyhedra gives results which are {\em in excellent
  agreement} with exact continuum values for these quantities. We have
further tested the performance of SURFGEN against Gaussian random
fields for which the analytic prediction for MFs is available and find
that our calculations reproduce the analytic results to a remarkable
precision. This gives us confidence that our ansatz is well suited for
application to large N-body simulations and three-dimensional galaxy
redshift surveys. Encouraged by the success on these preliminary
tests, we apply our method to simulations of three cosmological models
due to the Virgo consortium.

The rest of this paper is arranged as follows: In Section 2, we
provide a brief overview of Minkowski functionals and present an
algorithm which computes MFs for a triangulated polyhedral surface. In
Section 3, we describe the SURFGEN algorithm which triangulates a
general class of surfaces including iso-density surfaces describing
superclusters or voids.  In Section 4 we test our method against
standard eikonal surfaces $-$ spheres, ellipsoids, tori $-$ and
demonstrate the accuracy of our algorithm. Special boundary conditions
are needed to reproduce the analytic expressions for MFs of a Gaussian
random field. The related discussion along with the results for
Gaussian random fields is presented at the end of the paper in an
appendix. Section 5 is devoted to a detailed morphological study of
cosmological N-body simulations.  The models investigated are $\l$CDM,
$\T$CDM and SCDM at the present cosmological epoch.  Our conclusions
are presented in Section 6 which also discusses possible further
applications of SURFGEN and the MFs.

\section{Minkowski Functionals and Triangulated Surfaces}
\label{sec:MF}

The large scale structure of the universe can be studied on various
scales by considering the geometry and topology of excursion sets of
the density contrast $\delta({\bf x})$ defined as ${\cal E}^+_{TH} =
\lbrace {\bf x} |\delta({\bf x}) \geq \delta_{TH}\rbrace$, for
overdense regions (clusters, superclusters) and ${\cal E}^-_{TH} =
\lbrace {\bf x} |\delta({\bf x}) \leq \delta_{TH}\rbrace$, for
underdense regions (voids). By specifying a given density threshold
one effectively defines an iso-density surface over which the
Minkowski functionals should be evaluated.  Depending upon whether the
surface encloses overdense region of the space or underdense region,
the surface refers to a cluster/supercluster or a void.  Note that the
term `cluster' is used for any region in the excursion set ($\delta >
\delta_{TH}$) which in general is different from the astronomical
definition.  The terminology reflects the cluster analysis jargon.
The following four Minkowski functionals describe the morphological
properties of an iso-density surface in three dimensions
\footnote{There are ($n+1$) Minkowski functionals in $n$ dimensions
  each defined for an ($n-1$) dimensional hypersurface. Two
  dimensional MFs have been used to analyze the LCRS slices by
  Bharadwaj et al. (2000) and the anisotropy of the cosmic microwave
  background by Schmalzing \& Gorski (1998) and by Novikov, Feldman \&
  Shandarin (1999).}
%%%%
\begin{itemize}
\item {[1]}{\it Area} $S$ of the surface,
\item {[2]}{\it Volume} $V$ enclosed by the surface,
\item {[3]}{\it Integrated mean curvature} $C$ of the surface (or
  integrated extrinsic curvature), 
%%%%
\b
\label{eq:curv}
C = \frac{1}{2}\oint{\left({1\over R_1} + {1\over R_2}\right)dS}, 
\e
%%%%
where $R_1$ and $R_2$ are the principal radii of curvature at a given
point on the surface.
\item {[4]} {\it Integrated intrinsic (or Gaussian) curvature} $\chi$ of the
  surface -- also called the {\it Euler characteristic}
%%%%
  \b
\label{eq:euler} \chi = \frac{1}{2\pi}\oint{\left({1\over
      R_1R_2}\right)dS}.  
\e
%%%%
\end{itemize}
%%%%
A related measure of topology is the genus $G = 1 - \chi/2$.
\footnote{Genus measures the number of handles in excess of the number
  of isolated underdense regions (voids) that the surface of a cluster
  exhibits.}  Multiply connected surfaces have $G > 0$ while $G = 0$
  for a simply connected surface such as a sphere.  (The topological
  properties of all orientable surfaces are equivalent to those of a
  sphere with one or more `handles'. Thus a torus is homeomorphic to a
  sphere with one handle, while a pretzel is homeomorphic to a sphere
  with two handles etc.)  While the genus provides information about
  the connectivity of a surface, the remaining three MFs are sensitive
  to local surface deformations and hence characterize the geometry
  and shape of large scale structure at varying thresholds of the
  density \citep{sss98}.

In nature one seldom comes across surfaces that are perfectly smooth
and differentiable.  As a result expressions [1] - [4] which make
perfect sense for manifolds ${\cal C}^n$, $n \geq 2$ are woefully
inadequate when it comes to determining the Minkowski functionals for
real data sets which are grainy and quite often sparse.  Below we
describe how one can determine the Minkowski functionals (and derived
quantities) for surfaces derived from real data by interpolation.  The
interpolation scheme which describes iso-density contours as
triangulated surfaces will be presented in detail in the next Section.

We construct a polyhedral surface using an assembly of triangles in
which every triangle shares its sides with each of its three
neighboring triangles.
\begin{itemize}
\item The area of such a triangulated surface is 
\b 
S = \sum_{i=1}^{N_T}{S_i}, 
\e 
%%%%
where $S_i$ is the area of the $i-$th triangle and $N_T$ is the
total number of triangles which compose a given surface.
\item The volume enclosed by this polyhedral surface is the summed
  contribution from $N_T$ tetrahedra 
%%%%
\ber
\label{volume}
V &=& \sum_{i=1}^{N_T}{V_i},\nonumber\\
V_i &=& \frac{1}{3} S_i (n_j{\bar P}^j)_i.
\eer
%%%%%
Here $V_i$ is the volume of an individual tetrahedron whose base is a
triangle on the surface.  $(n_j{\bar P}^j)$ is the scalar product
between the outward pointing normal $\hat{n}$ to this triangle and the
mean position vector of the three triangle vertices, for which the
$j^{th}$ component is given by
%%%%%
\b
\label{eq:volume1}
{\bf{\bar P^j}} = \frac{1}{3}(P_1^j + P_2^j + P_3^j).  
\e 
%%%%%
The subscript $i$ in (\ref{volume}) refers to the $i-$th tetrahedron,
while the vectors $P_1, P_2, P_3$ in (\ref{eq:volume1}) define the
location of each of three triangle vertices defining the base of the
tetrahedron relative to an (arbitrarily chosen) origin
(Fig.~\ref{fig:volume}).  The vertices are ordered anticlockwise.
This ensures that the contribution to the volume from tetrahedra whose
base triangles lie on {\em opposite sides} of the origin add, while
the volumes of tetrahedra with base triangles lying on the {\em same
  side} of the origin subtract out.  Thus for both possibilities we
get the correct value for the enclosed volume (Fig.~\ref{fig:volume}).

%%%%%%%%%%%%%%%%%%%%
\begin{figure*}
\centering      
\centerline{
        \includegraphics[width=4in]{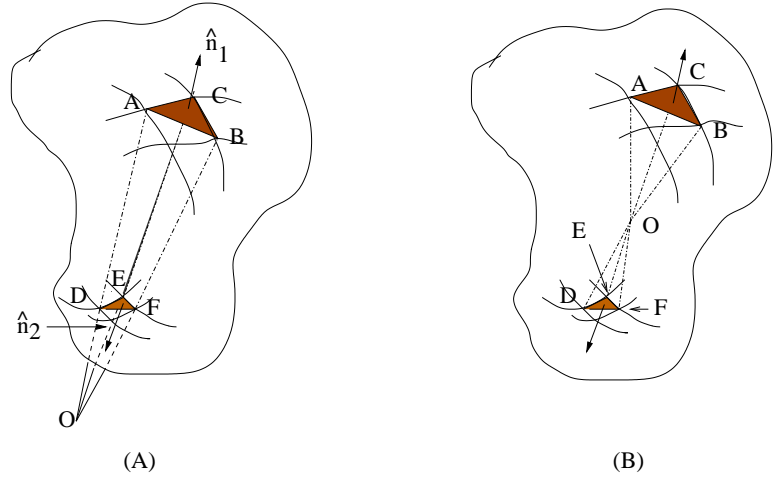}
        }       
\caption{{Volume enclosed by the surface is estimated by
    {\em vectorially} summing over the volumes of individual
    tetrahedra; each having its base at one of the triangles composing
    the surface; and its apex at an arbitrarily chosen point which we
    call the origin. Note that the origin can lie both within as well
    as outside the surface. The left panel shows how SURFGEN estimates
    the volume when the origin lies {\em outside} the surface.  In
    this case, the contribution (to the enclosed volume) from
    triangles falling in the same solid angle carries opposite sign
    for tetrahedra OABC and OFED. Clearly the volume of OFED must be
    subtracted from the volume OABC in order to give the true volume
    enclosed by the surface.  Anticlockwise orientation of the
    vertices used in determining the normals of the base triangles
    helps us bring this about. The right panel shows the second
    possibility for which the two triangles ACB and DEF form
    tetrahedra whose fourth (common) vertex lies {\em within} the
    surface. In this case, the contributions from ODEF and OACB add to
    give the total volume.}}
\label{fig:volume}
\end{figure*}
%%%%%%%%%%%%%%%%%%%%
\item {\em The extrinsic curvature of a triangulated surface is
    localized in the triangle edges.} As a result the integrated mean
  curvature $C$ is determined by the formula \b \label{eq:curvature} C
  = {1\over2}\sum_{i,j}{\ell_{ij}.\phi_{ij}}.\epsilon, \e where
  $\ell_{ij}$ is the edge common to adjacent triangles $i$ and $j$ and
  $\phi_{ij}$ is the angle between the normals to these triangles
  $\hat{n}_i$ \& $\hat{n}_j$.
%%%%
\b 
\cos\phi_{ij} = \hat{n}_i.\hat{n}_j.  
\e
%%%%
The summation in (\ref{eq:curvature}) is carried out over all pairs of
adjacent triangles. It should be noted that for a completely general
surface, the extrinsic curvature can be positive at some (convex)
points and negative at some other (concave) points on the surface. To
accommodate this fact one associates a number $\epsilon = \pm 1$ with
every triangle edge in (\ref{eq:curvature}). $\epsilon = 1$ if the
normals on adjacent triangles diverge away from the surface,
indicating a locally convex surface, while $\epsilon = -1$ if the
normals converge towards each other outside the surface, which is
indicative of a concave surface. In the former case the centre of
curvature of the surface lies within the surface-body whereas in the
latter case the centre of curvature lies outside of the surface-body
(Fig.~\ref{fig:curv}).

%%%%%%%%%%%%%%%%%%%%
\begin{figure*}
\centering
\centerline{
        \includegraphics[width=4in]{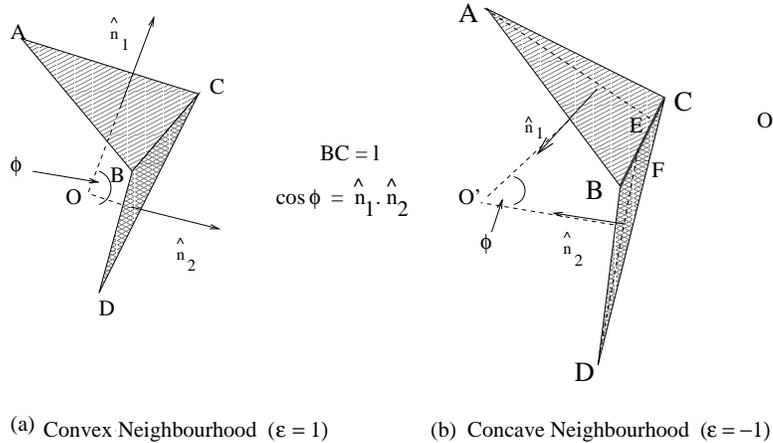}
        }       
\caption{{Illustrated here is the algorithm to compute the local
    contribution to the Integrated Mean Curvature: (a)$\Delta$ABC and
    $\Delta$CBD are two triangles sharing a side BC of length $\ell$.
    The normals $\hat{n}_1$ and $\hat{n}_2$ to the two triangles lie
    in the planes orthogonal to BC.  Their projection in the plane of
    the paper diverges from a point O {\it inside} the surface.  Thus
    the triangles correspond to a convex neighborhood and $\epsilon=1$
    in Eq.(\ref{eq:curvature}).  The angle between the two triangles
    is denoted as $\phi$. (b) Shown here are a similar pair of
    triangles, but with inverted sense of the normals, so that the
    sense of the surface is opposite to that shown in (a). We note
    that in the present case, the two normals converge {\it outside}
    the surface, so that their projections in the plane of the paper
    would meet at O', which lies outside the surface. Thus this
    represents a concave neighborhood, for which $\epsilon=-1$ in
    Eq.(\ref{eq:curvature}).  The angle $\phi$ represents the internal
    angle between $\hat{n}_1$ and $\hat{n}_2$.}}
\label{fig:curv}
\end{figure*}%fig2
%%%%%%%%%%%%%%%%%%%%

\item The genus of a triangulated closed polyhedral surface is given
  by the convenient expression
%%%%
\ber\label{eq:genus}
G &=& 1 - {\chi\over2}, \nonumber\\
\chi &=& N_T - N_E + N_V, 
\eer
%%%%
where $\chi$ is the Euler characteristic of the triangulated surface.
$N_T, N_E, N_V$ are, respectively, the total number of triangles,
triangle-edges, and triangle-vertices defining the surface.

\item As demonstrated in \citet{sss98,sss98b} ratio's of Minkowski
  functionals called ``Shapefinders'' provide us with an excellent
  measure of morphology.  Therefore, in addition to determining MFs we
  shall also determine the Shapefinders, $T$ (Thickness), $B$
  (Breadth) and $L$ (Length) which are defined as follows:
%%%%
\b \label{eq:s123}
T = {3V\over S}, \ \  B = {S\over C}, \ \ L = {C\over 4\pi(G+1)},
\e 
where $G = 0$ for simply connected surfaces and $G >$ 0 for multiply
connected regions.  The three Shapefinders (describing an object) have
dimensions of length and provide us with an estimate of the
`extension' of the object along each of the three spatial directions:
$T$ is the shortest and thus describes the characteristic thickness of
the object, $L$ is the longest and thus characterizes the length of
the object; $B$ is intermediate and can be associated with the breadth
of the object.  This simple interpretation obviously is relevant only
for fairly simple shapes.  For example, a triaxial ellipsoid has the
values of $T$, $B$ and $L$ closely related to the lengths of the three
principal axes: shortest, intermediate and the longest respectively.
It is worth stressing that $L$ defined by eq. (\ref{eq:s123})
quantifies the characteristic length between holes which should be
taken into account when interpreting the results (\eg Table
\ref{tab:top10s123} below).

\item An excellent indicator of `shape' is provided by the
  dimensionless Shapefinder statistic \b \label{eq:shapefinder} P =
  {B-T\over B+T}; F ={L-B\over L+B}, \e where $P$ and $F$ are measures
  of Planarity and Filamentarity respectively ($P, F \leq 1$).  A
  sphere has $P = F = 0$, an ideal filament has $P = 0, F = 1$ while
  $P = 1, F = 0$ for an ideal pancake. Other interesting shapes
  include `ribbons' for which $P \sim F \sim 1$.  When combined with
  the genus measure, the triplet $\lbrace P,F,G \rbrace$ provides an
  example of {\em shape-space} which incorporates information about
  topology as well as morphology of superclusters and
  voids.\footnote{Non-geometrical shape-statistics based on mass
    moments etc. can give misleading results when applied to large
    scale structure, as demonstrated in \citet{sss98b}.}

\end{itemize}

Having presented an ansatz to calculate MFs and Shapefinders on a
triangulated surface, we now discuss the {\em surface generating}
algorithm (SURFGEN) which creates triangulated surfaces corresponding
to iso-density contours evaluated at any desired threshold of the
density field.

\section{Surface Generating Algorithm (SURFGEN)}
SURFGEN is a versatile and powerful prescription which allows us to
generate and study surfaces whose physical origin can be quite varied
and different. Intensity surfaces (isophotes), isotherms and
iso-density surfaces constructed from 3D data provide examples of two
dimensional surfaces which can be generated and studied using SURFGEN
and the Minkowski functionals discussed in the previous
section\footnote{SURFGEN is a modified and extended version of the
  Marching Cubes Algorithm (MCA) which is used in the field of medical
  imaging to render high resolution images of internal organs by
  processing the output generated by X-ray tomography 
  \citep{mca}.}. The present paper is the first of a series in which
SURFGEN and MFs will be used in conjunction to make a detailed
morphological study of the supercluster-void network in the universe.
SURFGEN will be applied to a density field $\rho(\bf{r})$ defined on a
rectangular cubic lattice.

The SURFGEN algorithm is constructed as follows:

\begin{enumerate}
  
\item{} A point particle distribution (from N-body simulations, galaxy
  catalogues) is used to reconstruct the density field on the vertices
  of a cubic lattice via a cloud-in-cell (CIC) approach.  The lattice
  itself consists of a large number of closely packed elementary
  cubes. Any given cube is characterized by the value of the density
  at its eight vertices.
%\item{}  
% \item{} 
\item{} An appropriate density threshold $\rho_{\rm TH}$ is chosen and
  lattice points at this threshold are found using linear
  interpolation. The lattice-cells which involve such interpolated
  points along their sides are marked for triangulation which is to be
  done in the next step.
%The grid can be visualized as consisting of a large
%number of closely packed grid-cubes.  Any given grid-cube is

\item{} SURFGEN determines a closed polyhedral surface at 
the threshold $\rho_{TH}$ by {\em triangulating elementary cubes} 
while simultaneously moving
across the lattice. Triangulated elementary cubes are then put 
together to make up the full two-dimensional triangulated surface.

\end{enumerate}

The process of triangulation is based on the
following observation.

For any arbitrary threshold of the density, all elementary cubes of the density
field fall into one of the following three classes:

%%%%%%%%%%%%%%%
\begin{figure*}
\centering
\centerline{
  \includegraphics[width=4in]{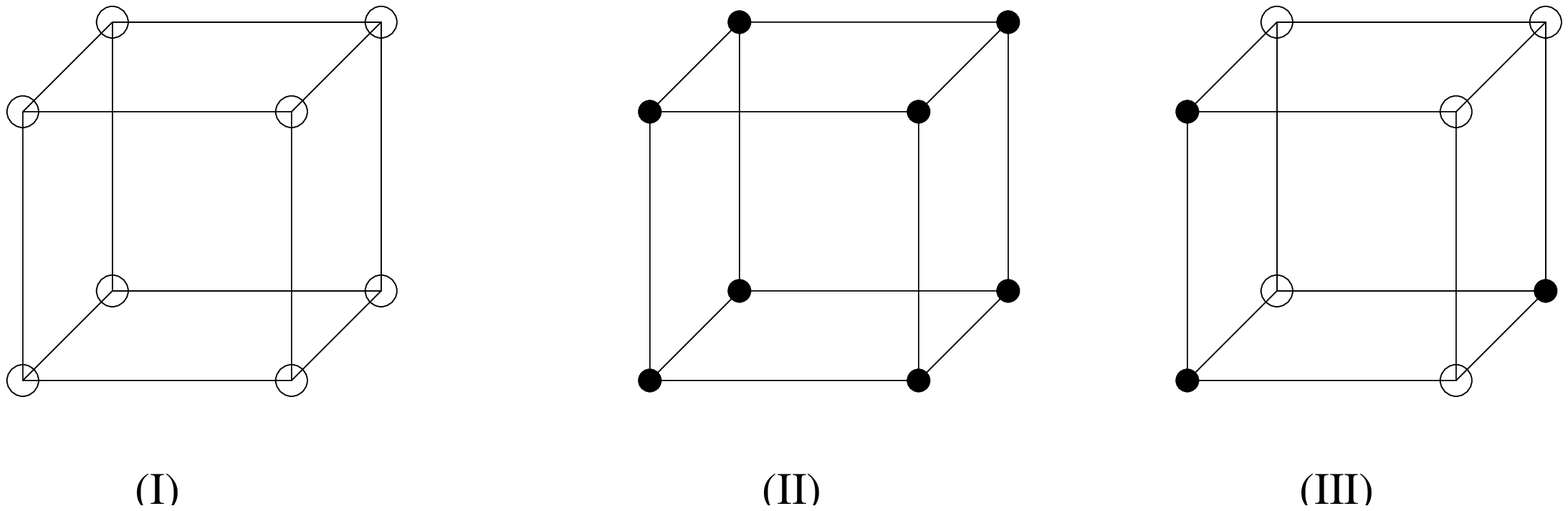} }
\caption{Given a density field defined on a lattice, the elementary 
lattice-cubes
  fall into one of the above three classes: (1) The cubes which are
  completely underdense, (2) the cubes which are completely overdense
  and finally (3) such cubes which have a few vertices overdense and
  the remaining vertices underdense. Cubes referring to the $3^{\rm
    rd}$ category are triangulated to model the surfaces.}
\label{fig:threetypes}
\end{figure*}
%%%%%%%%%%%%%%%%
\smallskip (i) those which have all eight vertices below the density
threshold ({\em underdense cubes} -- Fig.~\ref{fig:threetypes}$-$I),

(ii) those which have all eight vertices above the density threshold
({\em overdense cubes} -- Fig.~\ref{fig:threetypes}$-$II),

(iii) cubes having both overdense and underdense vertices ({\em
  surface cubes} -- Fig.~\ref{fig:threetypes}$-$III).

\smallskip

When modelling clusters and superclusters overdense cubes will be {\em
  enclosed} by our surface while underdense cubes will be excluded by
it. (For voids the situation will be reversed.)  Contour surfaces at a
prespecified density threshold will {\em lie entirely within surface
  cubes}. Thus the properties of {\em surface cubes} are vitally
important for this surface-reconstruction exercise.

We now describe how SURFGEN constructs a surface at the desired
threshold $\rho_{TH}$ by triangulating {\em surface cubes}.  We work
under the assumption that the underlying density field $\rho({\bf r})$
is continuous, so that in moving from an overdense site to an
underdense site (along a cube edge) we invariably encounter a point at
which $\rho = \rho_{TH}$.  SURFGEN classifies a given lattice-cube in
terms of the number of points (on the edges) where $\rho = \rho_{TH}$.
The precise location of these points (found by interpolation) tells us
exactly where our iso-density surface will intersect the cube.  Since
there are eight vertices and each vertex can either be overdense (1)
or underdense (0), the number of possible configurations of the cube
are $2^8 = 256$. Clearly the triangulation must be invariant upon
interchanging 1's with 0's, and this reduces the number of independent
configurations to 128.  Upon invoking rotational symmetry this number
further reduces to only 14. (For instance, although there are eight
cubes which have a single overdense vertex, any two members of this
family are related via the three dimensional rotation group. Therefore
a single scheme of triangulation suffices to describe all members of
this group.)
%%%%%%%%%%%%%%%%%
\begin{figure*}
\centering
\centerline{
        \includegraphics[width=5in]{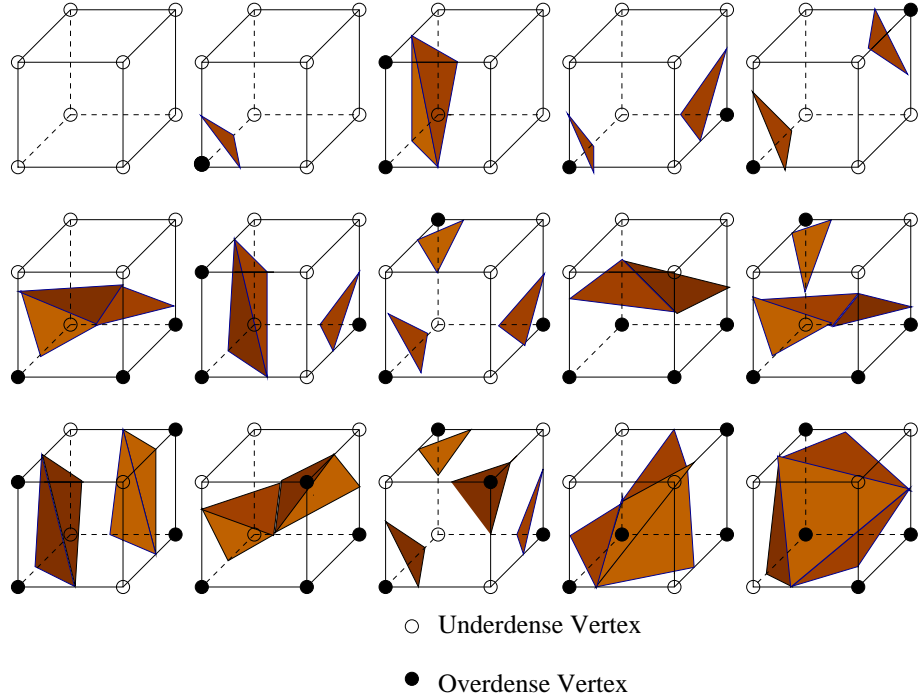}
        }
\caption{{Depending upon the value of density (or temperature, if isotherms
    are being constructed and photon counts, if isophotes are being
    modelled) at its vertices, a surface cube can be triangulated
    according to one of the fourteen independent configurations shown
    above. The black dots represent overdense vertices whereas
    underdense vertices are represented by open circles.}}
\label{fig:MCA}
\end{figure*}
%%%%%%%%%%%%%%%%%
These fourteen configurations of triangulated cubes are shown in
Fig.~\ref{fig:MCA} (also see \citet{mca}).

The surface intersecting a given {\em surface cube} forms quite
clearly only a portion of the full iso-density surface which we are
interested in constructing.  By explicitly demanding that the density
field be {\it continuous} across the iso-density surface one joins
triangles across the faces of neighboring cubes, thus constructing the
full continuous polyhedral surface.  This then is the basic
prescription which may be used to construct a closed triangulated
surface at any desired value of the density threshold.  However, a few
comments about triangulation are in order:
%%%%%%%%%%%%%%%%%%%
\begin{figure}
\bc
\resizebox{8.5cm}{!}{\includegraphics{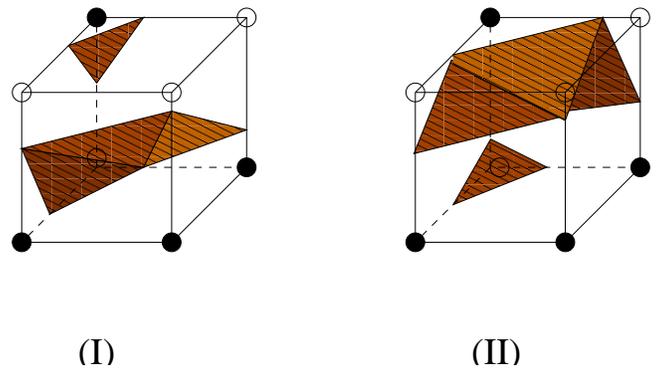}}
\caption{The cube shown here is the last
  in the middle row of Fig.4. It has four overdense and underdense
  vertices and, as shown above, it can be triangulated in two distinct
  ways.  Two other cubes (not shown) also allow for this possibility
  (of being triangulated in two distinct ways).  These are the last
  two cubes belonging to the last row of Fig.4. The original MCA
  algorithm does not mention this degeneracy. However, it can be shown
  that this degeneracy can be broken if we include information about
  neighboring cubes.  Requiring that the density field be continuous
  across the neighboring cubes breaks an occasional degeneracy which
  may arise, and allows one to triangulate surfaces unambiguously.}
\label{fig:degen}
\ec
\end{figure}
%%%%%%%%%%%%%%%%%%%
(i) the triangulation of the cubes having four contrasting vertices is
{\em not unique}. To illustrate this let us consider
Fig.~\ref{fig:degen} where the last cube belonging to the middle row
of Fig.~\ref{fig:MCA} is reproduced in the left panel.  We note that
this cube is symmetric with respect to an interchange of underdense
sites with overdense sites. As a result, we could just as well
triangulate it according to the prescription shown in the right panel
of Fig.~\ref{fig:degen}. This degeneracy in the way a given cube can
be triangulated is inherent also in the last two cubes of the last row
of Fig.~\ref{fig:MCA}. Noticeably, at all such places where we
encounter one of these cubes, one scheme of triangulation has to be
preferred over the other, otherwise the surface could become
discontinuous. Thus to create a unique triangulated surface one must
triangulate such cubes {\em in tandem} with their neighbors, ensuring
that the continuity of the surface is maintained.
%%%%%%%%%%%%%%%%%%%
\begin{figure}
\bc
\resizebox{8cm}{!}{\includegraphics{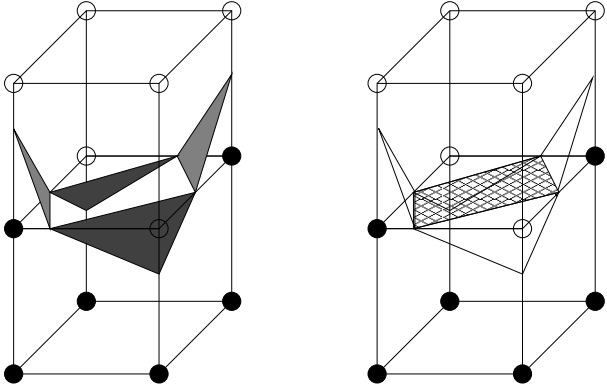}}
\caption{{Formation of a `hole' is illustrated. We `fill' this hole by
    a pair of triangles as shown in the right panel of the figure.
    Since the original MCA algorithm was meant strictly for
    visualization purposes, this problem appears to have been
    overlooked in Lorenson and Cline (1987).  We correct for this in
    our triangulation scheme by filling holes whenever they occur
    since its of prime importance to work with closed polyhedral
    surfaces when constructing MFs. }}
\label{fig:holeform}
\ec
\end{figure}
%%%%%%%%%%%%%%%%%%%
(ii) There are instances when two neighboring cubes have a density
configuration which does not lead to the complete closure of the given
surface. The resulting surface has a hole (Fig.~\ref{fig:holeform}a)
which must be filled by constructing two additional triangles which
close the surface (Fig.~\ref{fig:holeform}b).  The probability of hole
formation is finite every time the common face shared by two cubes
shows two vertices of type 1 along one diagonal and of type 0 along
the other diagonal. 

\begin{figure*}
\centering
\centerline{
\includegraphics[width=5in]{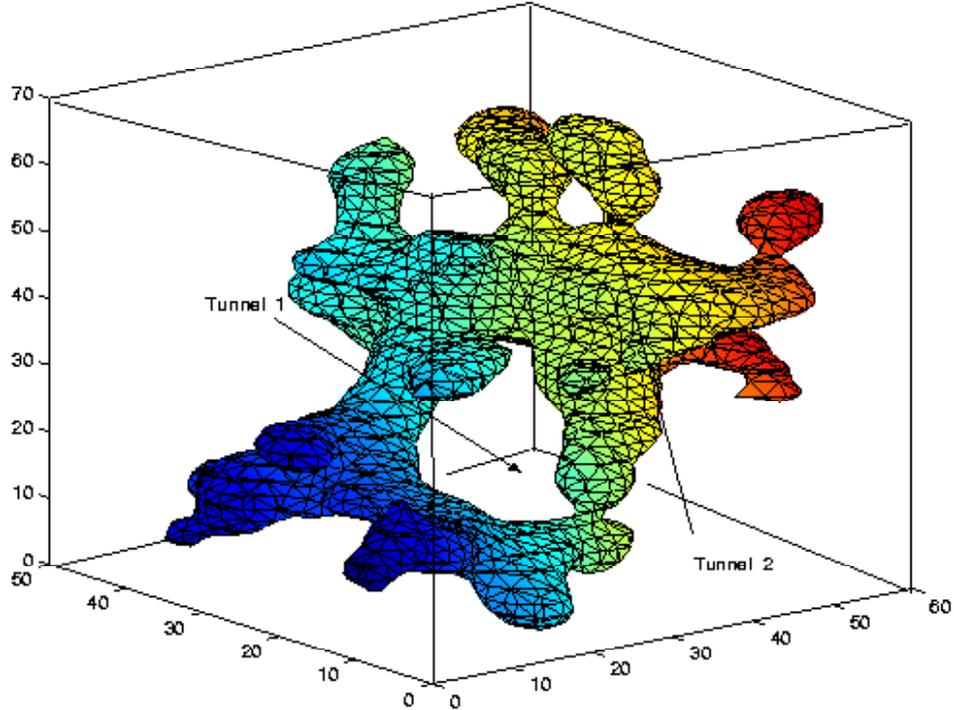}
}
\caption{{A cluster appearing in
the $\l$CDM simulation is triangulated using SURFGEN.  
The cluster is multiply connected with two tunnels and hence is described
by a genus value of two.}}
\label{fig:VSC}
\end{figure*}

Thus, in order to generate a closed polyhedral surface at a given
value of the density threshold $\rho = \rho_{TH}$ SURFGEN uses the 14
independent triangulations of an elementary lattice cube in
conjunction with an algorithm which ensures that the surface is
continuous.  We should emphasize that surfaces generated in this
manner {\em need not be simply connected} (see Fig. \ref{fig:VSC}).
Indeed iso-density surfaces constructed at moderate density thresholds
in N-body simulations as well as 3D galaxy surveys tend to show a
large positive value for the genus (Eq.\ref{eq:genus}).  This issue
will be discussed in detail in forthcoming paper(s). To enable an
online calculation of the MFs, our surface modelling scheme also
adheres to the following requirements:
\begin{itemize}
\item The vertices of all the triangles are stored in an anticlockwise order.
  This enforces a uniform prescription on normals which always point
  outward from the surface being modelled. Information regarding the
  directionality of normals is of great importance since it is used
  both for calculating the volume as well as the mean curvature.
\item For online calculation of the mean curvature, information
  concerning a given triangle must be supplemented with information
  concerning the triplet of triangles which are its neighbors. This
  allows us to unambiguously determine the contribution to the mean
  curvature from a local neighborhood.
\item The total number of triangles, triangle-edges and
  triangle-vertices of a triangulated surface are counted which enable
  us to determine its genus when surface construction is complete.
\end{itemize}
%
%%%%%%%%%%%%%%%%%%%%
%%%%%%%%%%%%%%%%%%%%

We might add that, although in this paper the density field is
reconstructed from N-body particle positions by means of a CIC
prescription, this is not a necessary prerequisite for SURFGEN which
is versatile enough to be used to triangulate iso-density contours of
fields reconstructed using more elaborate techniques such as Delauney
tesselations \citep{sw00,rien}, Wiener reconstruction or adaptive
smoothing. As it turns out, these schemes tend to maintain the
structural complexity in the system without diluting the patterns in
an isotropic manner. However, since SURFGEN constructs a triangulated
surface by `marching across neighboring lattice cubes' \citep{mca}
the density field must be specified on a cubic lattice for this
approach to give meaningful results.  \footnote{We should emphasize
  however that the ansatz developed in section \ref{sec:MF} for
  determining the Minkowski functionals for a triangulated surface can
  easily be adapted to other triangulation schemes such as those
  described in \citet{rourke98,rien}.}

To summarize, our surface modelling code incorporates the
triangulation of the entire set of 256 possible configurations of a
cube within a single scheme and makes possible online computation of
the area, the volume, the integrated mean curvature and the genus of a
triangulated surface.  The code has been tested on a variety of
standard density distributions and eikonal morphologies as well as on
Gaussian random fields.  In the next section, we present results based
on this analysis.

\section{Results for Standard Eikonal Surfaces}
In order to test the accuracy of our ansatz we generate triangulated
surfaces whose counterpart continuum surfaces have known (analytically
calculable) Minkowski functionals.

\subsection{Spherical structures}
In the first exercise, we consider spherically symmetric density
distributions and generate surfaces of constant density for a variety
of density-thresholds starting from a chosen maximum radius down to
grid-size.  A convenient density distribution for this exercise is
%%%%%%%%%%%%%%%%%%%%%%%%
\ber \label{eq:sphden}
\rho(i,j,k) = \left\{
\begin{array}{ll}
\frac {\rho_0}{R}, & (i,j,k) \ne (i_0, j_0, k_0),\\
\rho_0, & (i,j,k) = (i_0, j_0, k_0), 
\end{array}
\right.
\eer
%%%%%%%%%%%%%%%%%%%%%%%%
where R is the distance between (i,j,k) and the centre.
%%%%%%%%%%%%%%%%%%%%%%%%
\b \label{eq:sphdist}
R = \sqrt{(i-i_0)^2+(j-j_0)^2+(k-k_0)^2}.
\e
%%%%%%%%%%%%%%%%%%%%%%%%
Since the density field falls off as the inverse of the distance from
the centre at ($i_0,j_0,k_0$), any threshold $\rho_{TH}$ is associated
with a sphere of radius $R = {\rho_0\over\rho_{TH}}$. Thus larger
spheres correspond to surfaces of lower constant density in this
model. We assume $i_0 = j_0 = k_0 = 15$, for the centre of the sphere
in our numerical calculations.

Applying the Shapefinders (\ref{eq:s123}) \& (\ref{eq:shapefinder}) to
a sphere of radius $R$ one finds the simple result $T=B=L=R$ and $F =
P = 0$.
%%%%%%%%%%%%%%%%%
\begin{figure*}
\centering
\centerline{
  \includegraphics[width=4in]{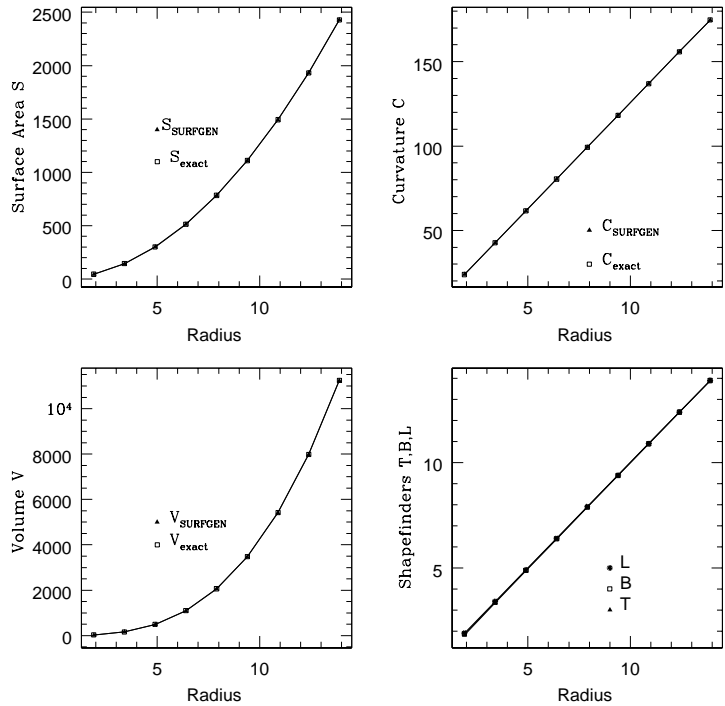}
  }       
\caption{{Minkowski Functionals and dimensional
    Shapefinders for a sphere. Note that the results for the Minkowski
    Functionals evaluated using our surface modelling scheme (SURFGEN)
    virtually coincide with the exact values.}}
\label{fig:Sph-Results}
\end{figure*}
%%%%%%%%%%%%%%%%%
Figure~\ref{fig:Sph-Results} compares the area, volume, mean curvature
and Shapefinders measured for a triangulated sphere against known
analytical values for these quantities. This figure clearly
demonstrates that exact values, and values computed using
triangulation, match exceedingly well down to the very lowest scale.

We should mention that the Genus evaluated using Eq.\ref{eq:genus} is
identically zero for spheres of all radii and for all possible
deformations of an ellipsoid (to be discussed next).  This provides an
excellent independent endorsement of our methodology by demonstrating
that our triangulations of a sphere and ellipsoid indeed result in
closed, continuous surfaces. 

\subsection{Triaxial ellipsoid}
A triaxial ellipsoid is an excellent shape with which to test a
morphological statistic. This is because, depending upon the relative
scales of the three axes, a triaxial ellipsoid can be oblate, prolate
or spherical.  We saw previously that MFs and Shapefinders give
extremely accurate results for spherical surfaces. We now demonstrate
that this remains true even for surfaces which are highly planar or
filamentary.

The parametric form of an ellipsoid having axes $a,$ $b,$ $c$ and
volume $V = \frac{4\pi}{3}abc$ is
\begin{equation}
{\bf{r}} = a(\sin{\theta}\cos{\phi})\hat{x} +
b(\sin{\theta}\sin{\phi})\hat{y}
+ c(\cos{\theta})\hat{z}
\end{equation}
where $0 \le \phi \le 2\pi$, $0 \le \theta \le \pi$.  For the purposes
of our study, we systematically deform a triaxial ellipsoid and study
how its shape evolves in the process. When shrunk along a single axis
a triaxial ellipsoid becomes planar.  Simultaneous shrinking along a
second axis makes it cigar-like or filamentary.

\subsubsection{Oblate spheroids}
We first study the accuracy of our triangulation scheme for planar
configurations by considering oblate deformations of an ellipsoid. In
this case two axes $a$ and $b$ are held fixed ($a\sim b$), while the
third axis $c$ is slowly shrunk leading to an increasingly planar surface.
%%%%%%%%%%%%%%%%%
\begin{figure*}
\centering
\centerline{
  \includegraphics[scale=0.9,trim=20 200 5 200,clip]{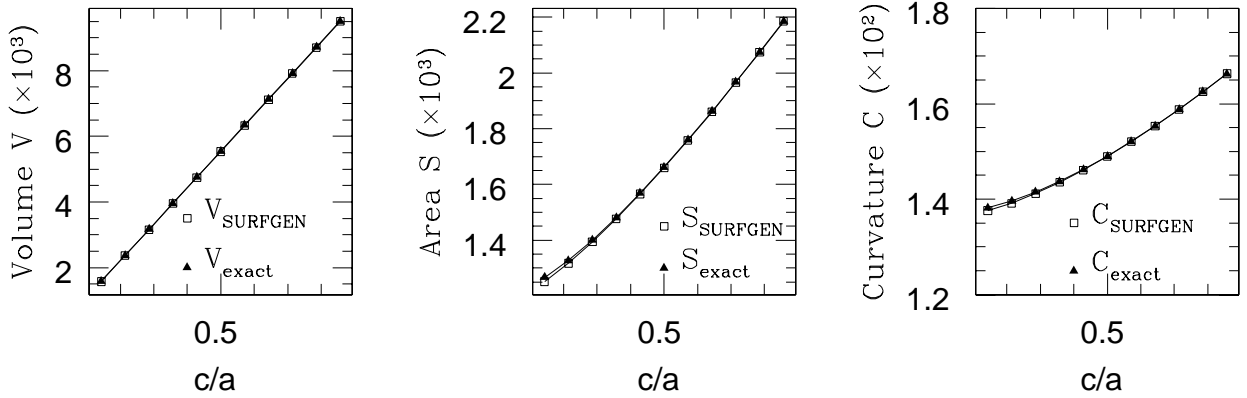}
        }
\caption{Values of Minkowski functionals determined by the surface modelling 
  scheme SURFGEN are shown along with their exact values for oblate
  deformations of an ellipsoid. Note that for $c/a > 0.2$, values for
  Minkowski Functionals evaluated using SURFGEN virtually
  coincide with the exact results.}
\label{fig:planmf}
\end{figure*}
%%%%%%%%%%%%%%%%%
%%%%%%%%%%%%%%%%%
\begin{figure*}
\centering
\centerline{
  \includegraphics[scale=0.8,trim=20 10 5 150,clip]{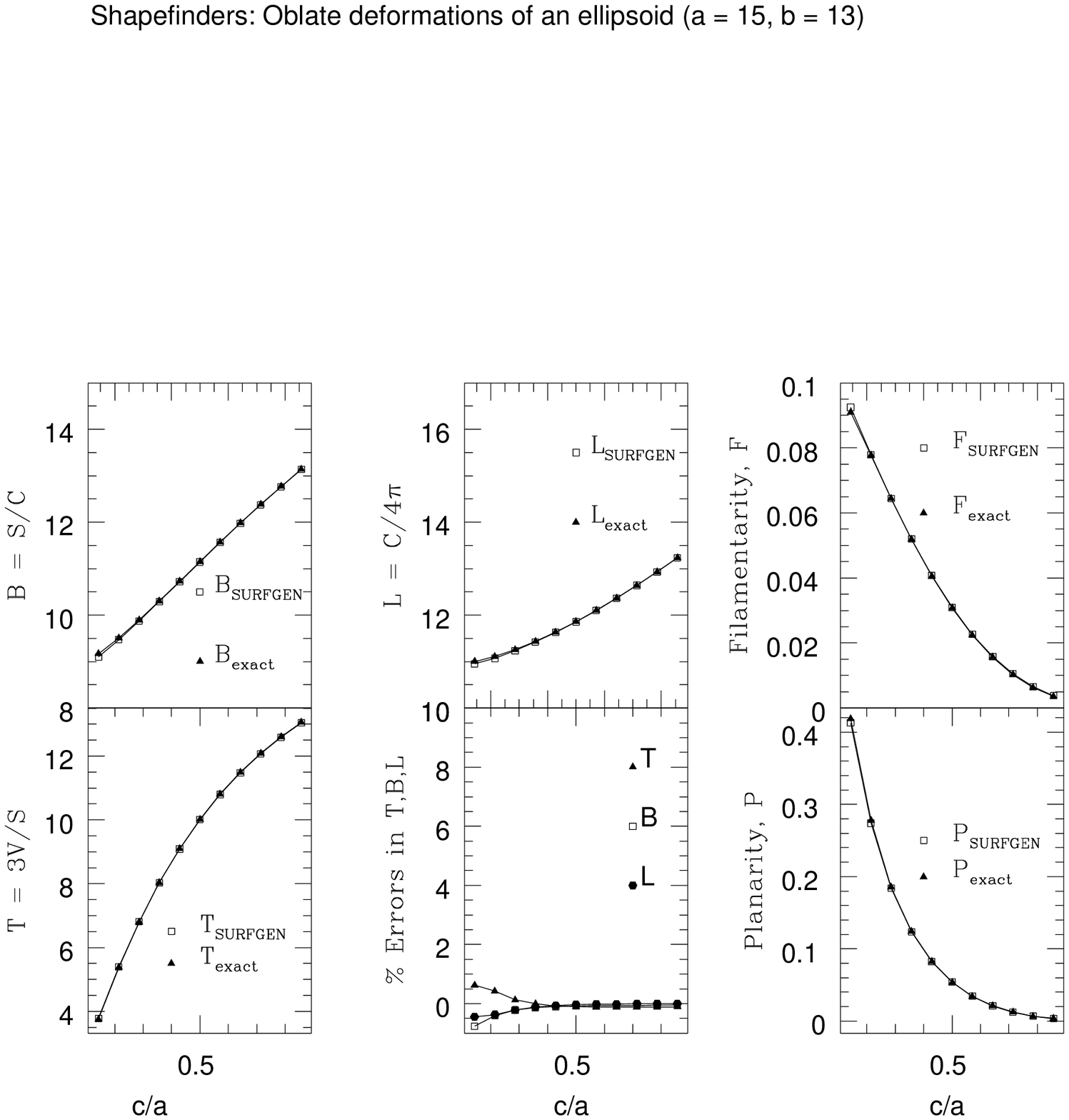}
        }
\caption{Values of Shapefinders derived using the surface modelling
  scheme SURFGEN are shown along with their exact values for oblate
  deformations of an ellipsoid.  The lower middle panel shows
  percentage errors in the estimation of these quantities. We note
  that the errors are all in the range of $\pm$0.9$\%$.  The right
  panels show the evolution of planarity and filamentarity with
  continuous shrinking of c-axis of the ellipsoid. Note that planarity
  increases from its low initial value whereas there is only a
  marginal increase in filamentarity.}
\label{fig:plans123}
\end{figure*}
%%%%%%%%%%%%%%%%%
Our results for this case are compiled in two figures.
Figure~\ref{fig:planmf} shows MFs as they evolve with the
dimensionless variable $c/a$. (The exact results for the Minkowski
functionals which we quote are based on the analytical expressions for
MFs given in \citet{sss98}, we refer the reader to that paper for more
details.)  Figure~\ref{fig:planmf} clearly demonstrates that the
values of MFs obtained using triangulation match the exact values to a
remarkable degree of accuracy. Indeed, for a wide range in $c/a$ the
two distinctly different calculational algorithms give virtually
indistinguishable results, thereby indicating that the triangulation
ansatz is, for all practical purposes, exact!

Figure~\ref{fig:plans123} shows the evolution of all the
Shapefinders, i.e., $T,B,L,P$ and $F$ together with the percentage
errors in the estimation of dimensional Shapefinders. We find that $L$
is estimated to greatest accuracy with maximum error of $\sim 0.4\%$,
while $T$ and $B$ can be determined to an accuracy of $\sim 0.8\%$.
We further note (right panels of Figure~\ref{fig:plans123}) that
Planarity grows from an initially low value $\sim 0.0$ to a large
final value $\sim 0.4$ as the ellipsoid becomes increasingly oblate.
Filamentarity, on the other hand remains small at $\sim 0.09$. Both
Filamentarity and Planarity are determined to great accuracy by
SURFGEN.

\subsubsection{Prolate spheroids}
Next we study prolate deformations of our oblate ellipsoid. We start
with $b \simeq a,$ $c \ll a,$ and shrink the second axis $b$ while
keeping $a$ and $c$ fixed, so that finally $c \sim b \ll a$ and our
initially oblate ellipsoid becomes prolate. Our results are again
summarized in two figures.

%%%%%%%%%%%%%%%%%
\begin{figure*}
\centering
\centerline{
  \includegraphics[scale=0.9,trim=20 200 5 200,clip]{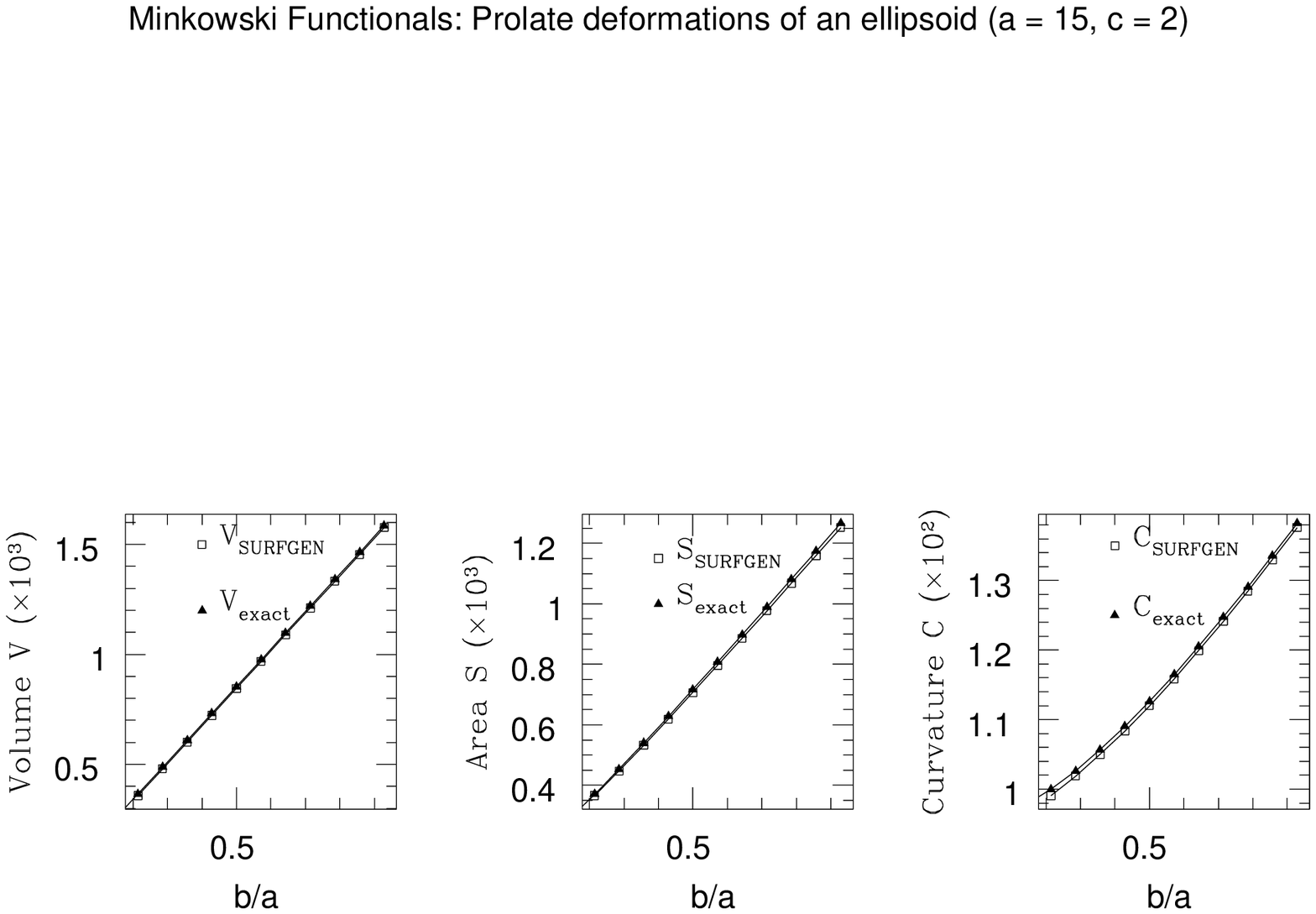}
        }
\caption{Values of Minkowski functionals determined by the surface modelling
  scheme SURFGEN are shown along with their exact values for prolate
  deformations of an ellipsoid.  Note that for $b/a \ge 0.3$, the
  results for the Minkowski Functionals evaluated using SURFGEN
  virtually coincide with the exact values.}
\label{fig:flmf}
\end{figure*}
%%%%%%%%%%%%%%%%%

Figure~\ref{fig:flmf} reveals very good agreement between measured and
true values of MFs, with the former tending to be slightly smaller
than the latter.

%%%%%%%%%%%%%%%%%
\begin{figure*}
\centering
\centerline{
  \includegraphics[scale=0.8,trim=20 10 5 150,clip]{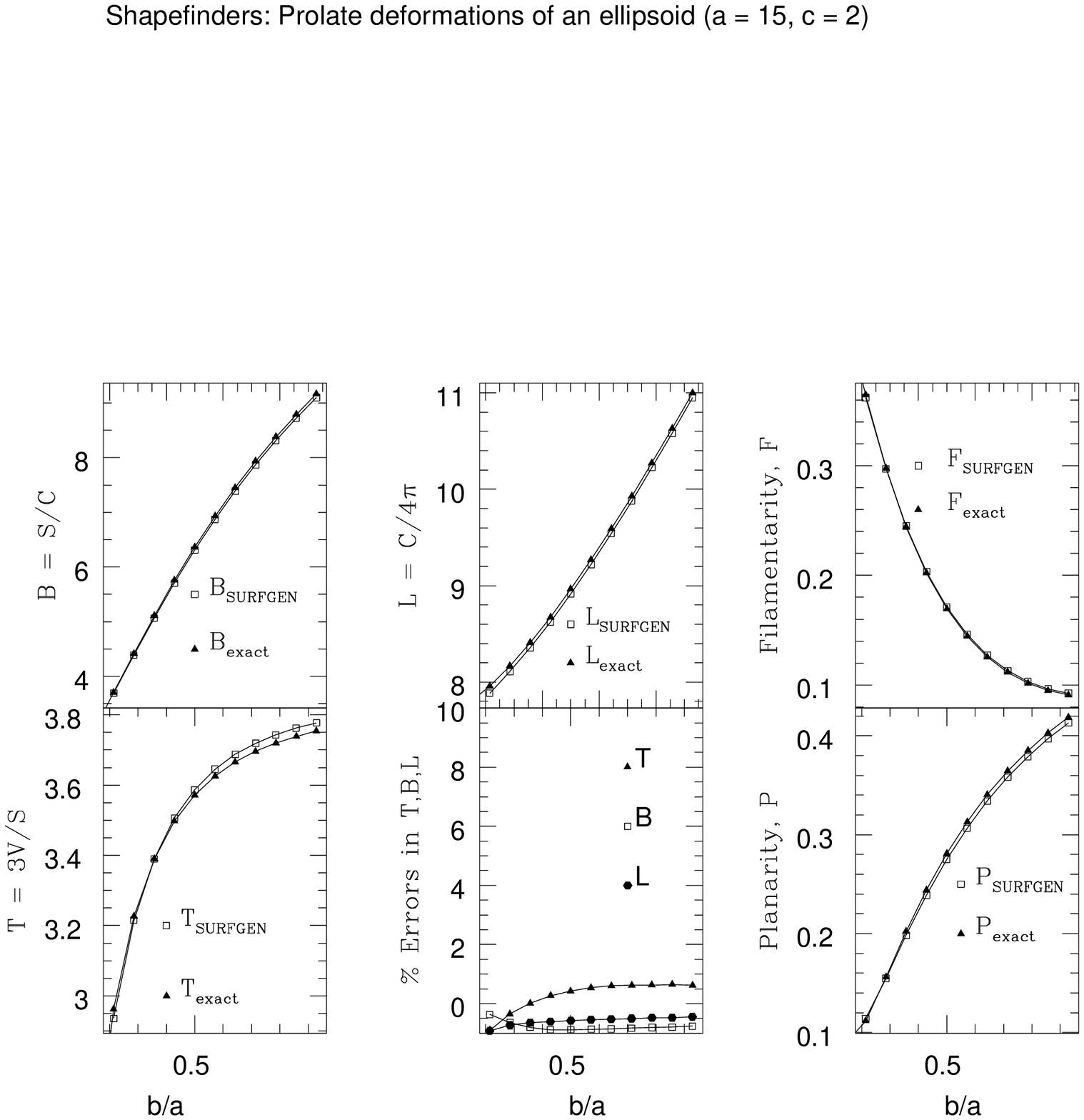}
        }
\caption{Values of the dimensional Shapefinders derived from the surface 
  modelling scheme SURFGEN are shown along with their exact values for
  prolate deformations of an ellipsoid.  The lower middle panel shows
  percentage errors in the estimation of these quantities. We note
  that the error in the estimation of all the quantities lies within
  +1$\%$ to -2$\%$. The right panels show the evolution of planarity and
  filamentarity with continous shrinking of the b-axis. Note that the
  planarity decreases from its high initial value and the filamentarity
  increases steadily.}
\label{fig:fls123}
\end{figure*}
%%%%%%%%%%%%%%%%%
Turning to the Shapefinders we find that, with the possible exception
of extremely prolate figures, the Shapefinders are remarkably well
determined. Indeed, even the extremely prolate ellipsoid with axis
ratio $b/a < 0.2$ has a largest error in $T$ of only $\sim 1.7\%$
while errors in $B$ and $L$ never exceed $\sim 1\%$ (Fig.
~\ref{fig:fls123}).
%%%%%%%%%%%%%%%%%
% \begin{figure*}
% \centering
% \centerline{
%         \includegraphics[width=4in]{filament_pf.eps}
%         }
% \caption{Values of dimensionless Shapefinders $P$ and $F$ derived using the 
%   surface modelling scheme SURFGEN are shown along with their exact
%   values for prolate deformations of an ellipsoid.}
% \label{fig:flpf}
% \end{figure*}
%%%%%%%%%%%%%%%%%
This figure also shows the evolution of planarity ($P$) and
filamentarity ($F$) as our ellipsoid becomes increasingly more
prolate.  In keeping with our expectation $F$ steadily increases from
a small initial value $\sim 0.1$ to $\sim 0.44$.  Planarity drops from
its large initial value to a small final value $P \sim 0.1$ and is
slightly underestimated for $b/a \ggeq 0.3$.  We therefore conclude
that both $P$ and $F$ are determined by SURFGEN to a sufficient
accuracy for prolate ellipsoidal figures.

\subsection{The Torus and its deformations}

Next we extend our analysis to manifolds which are multiply connected
by considering the deformations of a torus which has an elliptical
cross section and which we shall refer to as an elliptical torus.  A
torus is an important surface on which to test SURFGEN for two
reasons: it is multiply connected and, unlike an ellipsoid, it
contains regions which are convex (on its outside) as well as concave
(on its inside).  The elliptical torus can be described by three
parameters $a,b,c$. In this case the elliptical toroidal tube has
diameter $2\pi b$ and $a$ \& $c$ are its radii of curvature in two
mutually orthogonal directions. The elliptical torus reduces to the
more familiar circular torus when $a = c$. We choose to work with the
elliptical torus because changing the values of $a,b,c$ can give rise
to a large variety of surfaces all of which (by definition) are
multiply connected but which have very different shapes. Thus our
surface modelling scheme SURFGEN can be put to a rigorous test.  The
parametric form for the elliptic torus is
\b {\bf r} = (b + c\sin\phi)\cos\theta\hat{x} + (b +
c\sin\phi)\sin\theta\hat{y} + a\cos\phi \hat{z}, \e
where $a,c<b$, $0\le\phi<2\pi$ and $\theta<2\pi$.  We shall compare
results due to SURFGEN with the exact analytical results for four
kinds of tori - a nearly spheroidal torus, a ribbon, a pancake, and a
filament.
%%%%%%%%%
% Table-1 for Tori of various dimensions:
%%%%%%%%%
\begin{table*}
\centering
\caption{Minkowski Functionals for some extreme deformations of an
    elliptical torus. ${\Delta C\over C}, {\Delta S\over S}$ and 
${\Delta V\over V}$ give the percentage error in the determination of the 
Minkwoski functionals (MFs) using SURFGEN. An accuracy of better than 1\% 
is achieved for the three MFs: Curvature (C), Surface area (A) and 
Volume (V), while the genus is determined exactly.}
% \tablecolumns{9}
\begin{center}
\begin{tabular}{lllllllll} \hline
\multicolumn{1}{l} {$b,a,c$} &
\multicolumn{1}{l} {Surface} &
\multicolumn{1}{l} {C} &
\multicolumn{1}{l} {${\Delta C\over C}\%$} & 
\multicolumn{1}{l} {S} & 
\multicolumn{1}{l} {${\Delta S\over S}\%$} &
\multicolumn{1}{l} {V} & 
\multicolumn{1}{l} {${\Delta V\over V}\%$} & 
\multicolumn{1}{l} {Genus} \\
\hline
(40.0,37.9,37.9) & Sphere-with & $7.93\times10^2$ & $+0.47$ & $5.98\times10^4$ & $-0.01$ &  $1.13\times10^6$ & $-0.001$ & 1 \\
  & -hole & & & & & & & \\
  (140.0,19.9,1.99) & Ribbon & $2.8\times10^3$ & $+0.03$ & $7.1\times10^4$ & $-0.42$ & $1.1\times10^5$ & $-0.830$ & 1 \\
  (60.0,57.9,3.86) & Pancake & $1.2\times10^3$ & $+0.40$ & $8.8\times10^4$ & $-0.18$ & $2.7\times10^5$ & $+0.590$ & 1 \\
  $(50.02,3.52,3.52)$ & Filament & $9.92\times10^2$ & $+0.52$ & $6.95\times10^3$ & $+0.18$ & $1.21\times10^4$ & $-0.580$ & 1 \\
%  \enddata
\hline
\label{tab:mf}
\end{tabular}
\end{center}
\end{table*}
%%%%%%%%%
% Table-2 for Tori of various dimensions:
%%%%%%%%%
\begin{table*}
\centering
\caption{SURFGEN determined values of the Shapefinders $T$ (Thickness), $B$
  (Breadth), $L$ (Length), P (Planarity) and F (Filamentarity) describing 
extreme deformations of an elliptical torus. Also given alongside are the 
percentage errors in their estimation.}
\begin{center}
\begin{tabular}{llllllllllll} \hline
\multicolumn{1}{l} {Surface} &
\multicolumn{1}{l} {$T$} &
\multicolumn{1}{l} {${\Delta T\over T} \%$} & 
\multicolumn{1}{l} {$B$} &
\multicolumn{1}{l} {${\Delta B\over B} \%$} & 
\multicolumn{1}{l} {$L$} &
\multicolumn{1}{l} {${\Delta L\over L} \%$} & 
\multicolumn{1}{l} {P} &
\multicolumn{1}{l} {${\Delta P\over P} \%$} & 
\multicolumn{1}{l} {F} &
\multicolumn{1}{l} {${\Delta F\over F} \%$} & 
\multicolumn{1}{l} {Genus} \\
\hline
Sphere-with & 56.86 & 0.02 & 75.44 & $-$0.48 & 31.57 & 0.48 & 0.140 & $-$1.41 & $-0.410$ & 0.95 & 1 \\
-hole & & & & & & & & & & & \\
Ribbon & 4.67 & 1.08 & 25.63 & $-$0.47 & 109.96 & 0.03 & 0.692 & $-$0.57 & 0.622 & 0.32 & 1 \\
Pancake & 9.09 & 0.74 & 74.19 & $-$0.59 & 47.11 & 0.41 & 0.782 & $-$0.26 & $-$0.223 & 2.19 & 1 \\
Filament & 5.23 & $-$0.95 & 7.01 & $-$0.43 & 39.49 & 0.46 & 0.145 & 1.40 & 0.698 & 0.28 & 1 \\
\hline
\label{tab:s123}
\end{tabular}
\end{center}
\end{table*}
%%%%%
% Table-3 for Tori of various dimensions:
%%%%%
Tables \ref{tab:mf} and \ref{tab:s123} refer to these
four surfaces; relevant figures can be found in \citet{sss98}.
Table~\ref{tab:mf} shows the estimated values of the Minkowski
Functionals and the percentage error in their estimation.
Table~\ref{tab:s123} shows all the Shapefinders for these
surfaces\footnote{It should be noted that the definition of the third
  Shapefinder $L$ is modified from that reported in \citet{sss98}.
  New definition of $L$ is as given in Eq.\ref{eq:s123}. The newly
  adopted definition of $L$ incorporates treatment of simply
  connected and multiply connected surfaces in within a single
  scheme.}.  This table also shows the percentage errors in the
estimation of the Shapefinders.

We should emphasize that the genus estimated by SURFGEN for all these
deformations has the correct value of unity.  This gives an important
independent check on the self-consistency of the triangulation scheme
since, the absence of even a single triangle out of several thousand,
would create an artificial `hole' in our surface and give the wrong
value for the genus -- a situation that has {\em never} occurred for
any of the several dozen deformations of either the ellipsoid or the
torus.

Our results clearly demonstrate that the surface modelling scheme
SURFGEN provides values for the Minkowski functionals which are in
excellent agreement with exact analytical formulae.  We have further
tested SURFGEN against a set of Gaussian random field samples. Owing
to a subtle issue of the boundary conditions involved in the
corresponding test, we present the corresponding results in an
appendix at the end of this paper. For the time being, it suffices to
make a note that our ansatz for studying iso-density contours and
their morphology works exceedingly well both for simply connected as
well as multiply connected surfaces and on Gaussian random fields.

In the next section, we apply SURFGEN to cosmological N-body
simulations and give an example of the study of the morphology in three
models of structure formation: $\Lambda$CDM, $\tau$CDM and SCDM.

\section{Applications: Morphology of the Virgo Simulations}
As a first application of the SURFGEN code, we report here a
morphological study of cosmological simulations. We use dark matter
distributions of three cosmological models simulated by the Virgo
consortium. The models are (1) a flat model with $\Omega_0$ = 0.3
($\Lambda$CDM), and two models with $\Omega$ = 1, namely (2) one with
the standard CDM power spectrum (SCDM) and (3) a flat model with the
same power spectrum as the $\Omega_0$ = 0.3 $\L$CDM model (also referred to
as $\tau$CDM). The shape parameter $\Gamma$ (=$\Omega_0$h) for SCDM is
0.5, whereas for the other two models $\Gamma = 0.21$. Amplitude of
the power spectrum in all models is set so as to reproduce the
observed abundance of rich galaxy clusters at the present epoch. A
detailed discussion of the cosmological parameters and simulations can
be found in \citet{jenkcdm}.

The data consist of $256^3$ particles in a box of size 239.5
$h^{-1}$Mpc. In order to carry out a detailed morphological study, we
first need to smooth the data and recover the underlying density
field. The issue of density field reconstruction has been well
explored by several groups studying the topology of the large scale
structure. Here we follow the smoothing technique used by
\citet{springtop} which they adopted for their preliminary topological
analysis of the Virgo simulations. We fit a $128^3$ grid on to the
box. Thus, the size of each cell is 1.875 $h^{-1}$Mpc. Since there are
on an average 8 particles per cell, the smoothing is not too sensitive
to shot-noise.  The samples are fairly rich and we can smooth the
fields in two steps.  In the first, we apply a Cloud in Cell (CIC)
smoothing technique to construct a density field on the grid.  Next we
smooth this field with a Gaussian kernel which offers us an extra
smoothing length-scale.  There are various criteria to fix this
smoothing scale which depend on the relative sparseness/richness of
the samples. To avoid discreteness effect (so that there are no
regions in the survey where density is undefined), the simplest
criteria for sparse samples has been to set $\lambda \ge 2.5\ell_g$,
where $\ell_g$ is the grid-size.  There are other sophisticated
criteria to deal with sparse samples which are tested and reported in
literature, which need not be reiterated here on account of our
samples being rich. We are left with two choices; either to smooth the
fields at a scale comparable to $r_0$, the correlation scale, or to
probe the morphology at smaller length scales with an interest to
study the regularity in the behaviour of the MFs; of which the latter
is more promising. The average interparticle distance in the present
case is $\sim 0.94$ h$^{-1}$Mpc, which is far outweighed by the
grid-size and the correlation length. Hence, we adopt a modest scale
of smoothing, $\lambda = 2$ h$^{-1}$ Mpc, which is small compared to
usual standards as well as enough to provide the necessary smoothing.
The Gaussian kernel for smoothing that we adopt here is \footnote{
  Since the Gaussian kernel that we use is radially symmetric, it
  could, in principle, diminish the true extent of anisotropy in
  filaments and pancakes.  This effect could be minimized by
  considering anisotropic kernels and/or smoothing techniques based on
  the wavelet transform.  An even more ambitious approach is to
  reconstruct density fields using Delaunay tesselations
  \citep{rien}. Density fields reconstructed in this manner appear to
  preserve anisotropic features and may therefore have some advantage
  over conventional `cloud-in-cell' techniques followed by an
  isotropic smoothing \citep{sw00}. Thus, as far as the goal is to
  utilize the geometry of the patterns to discriminate between the
  models, such smoothing schemes should prove more powerful. We hope
  to return to these issues in forthcoming paper(s).}
%%%
\b W(r) = {1\over\pi^{3\over2}\lambda^3}
exp{\left(-{r^2\over\lambda^2}\right)}.  
\e 
%%%

In this section we first study the global MFs for all the three models.
Next we investigate the morphology of the percolating supercluster
network.  This is followed by a statistical study of the morphology of
smaller structures. We summarize our results in the next section.
\subsection{Global Minkowski Functionals}
We scan the density fields at 100 values of the density threshold
$\rho_{TH}$, all equispaced in the filling factor 
%%%
\b 
FF_{\rm V} = \int \Theta(\rho-\rho_{\rm TH}) d^3x, 
\e 
%%%
where $\Theta(x)$ is the Heaviside Theta function. $FF_{\rm V}$
measures the volume-fraction in regions which satisfy the `cluster'
criterion $\rho_{\rm cluster} \geq \rho_{\rm TH}$ at a given density
threshold $\rho_{\rm TH}$. In the following, we use $FF_{\rm V}$ as a
parameter to label the density contours.

At each level of the density field (labelled by $FF_{\rm V}$), we 
construct a catalogue of clusters (overdense regions) based on a
Friends of Friends (FOF) algorithm.\footnote{The clusters/superclusters
discussed in the present paper are defined as connected overdense regions lying
above a prescribed density threshold. Due to the large smoothing scale adopted,
the overdensity in clusters ranges from $\delta \sim 1$ to $\delta \sim 10$,
which makes them more extended ($ \ggeq $ few Mpc) and less dense than the
galaxy clusters in (for instance) the Abell catalogue.}
Next we (i) run the SURFGEN code on each of these clusters to model
surfaces for each of them and (ii) determine the Minkowski Functionals
(MFs) for all clusters at the given threshold (these are referred to
in the literature as partial MFs). Global MFs are partial MFs summed
over all clusters.  Thus, at each level of the density, we first
compute the partial MFs and then the global MFs. Our plots of global
MFs as functions of $FF_{\rm V}$
%%%%%%%%%%%%%%%%%%%%
\begin{figure*}
\centering      
\centerline{
  \includegraphics[width=5in]{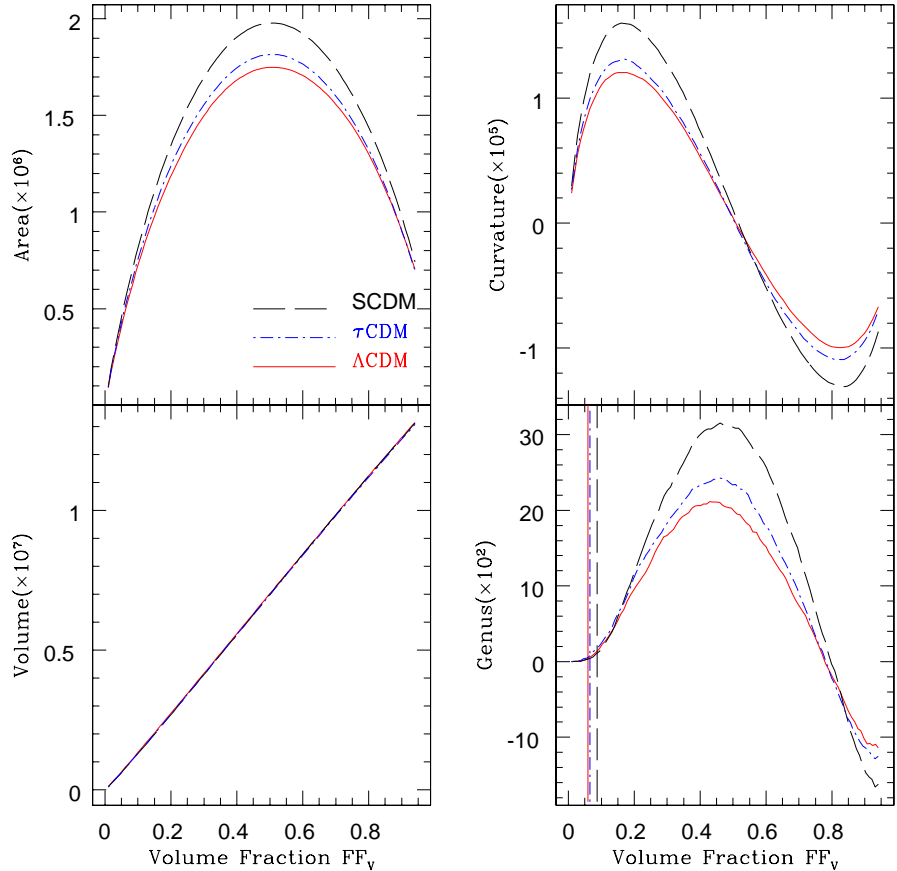}
  }       
\caption{{Global Minkowski functionals for $\Lambda$CDM, $\T$CDM and 
    SCDM are shown as functions of the volume fraction ($\equiv$
    volume filling factor, $FF_{\rm V}$).  The volume-curve is the
    same for all models and grows linearly with the volume fraction;
    this is simply a restatement of the definition of `filling
    factor'. Notice that the amplitude of the remaining three MFs
    (Area, Curvature and Genus) is substantially greater in SCDM than
    in $\l$CDM; with $\T$CDM falling between the two. This could mean
    that large scale structure is much more `spongy' in SCDM with
    percolating structures in this model showing many more `holes' (or
    tunnels) and resulting in a large value for the genus.  A relative
    shift towards left, of the position of the peak of the genus curve
    in $\L$CDM from $FF_V=$0.5 is indicative of the bubble shift,
    implying more amount of clumpyness in this model compared to other
    two models. It should be noted that for a Gaussian random field, the
    peak occurs at $FF_{\rm V} =$ 0.5. Gravitational clustering
    modulates the genus curve by lowering its peak and producing a
    shift to the left (the `bubble shift') or to right (the `meatball
    shift') depending upon the other model parameters. The relatively
    large value of the surface area and the curvature may indicate
    that moderately overdense superclusters have many more `twists and
    turns' in SCDM than in either $\tau$CDM or $\l$CDM, at identical
    values of the filling factor.  In all models the genus curve has a
    large negative value at large values of the volume fraction.  The
    reason for this is that the percolating supercluster, at extremely
    low density thresholds (high $FF_{\rm V}$), occupies most of the
    volume. Voids exist as small isolated bubbles in this vast
    supercluster and lead to a large negative genus whose value is of
    the same order as the total number of voids.  In the opposite
    case, at very high density thresholds associated with small
    filling fractions, we probe the morphology of isolated and simply
    connected clusters.  Since the genus for these clusters vanishes,
    the global topology of large scale structure at high thresholds of
    the density generically approaches zero for all the models.  The
    vertical lines in the lower right panel show the percolation
    threshold in SCDM (dashed), $\T$CDM (dot-dashed), and $\l$CDM
    (solid) which occurs at intermediate density thresholds
    corresponding to density contrast $\delta \sim 1$.  It is clear
    from this figure that the percolating supercluster has a
    relatively simple topology at the onset of percolation.  The rapid
    increase in genus value as one moves to lower density thresholds
    (larger $FF_{\rm V}$) reflects the progressive increase in the
    `sponge-like' topology of the percolating supercluster which is
    more marked in SCDM than in $\l$CDM.}}
\label{fig:globalff}
\end{figure*}
%%%%%%%%%%%%%%%%%%%
\begin{figure*}
\centering
\centerline{
  \includegraphics[width=5in]{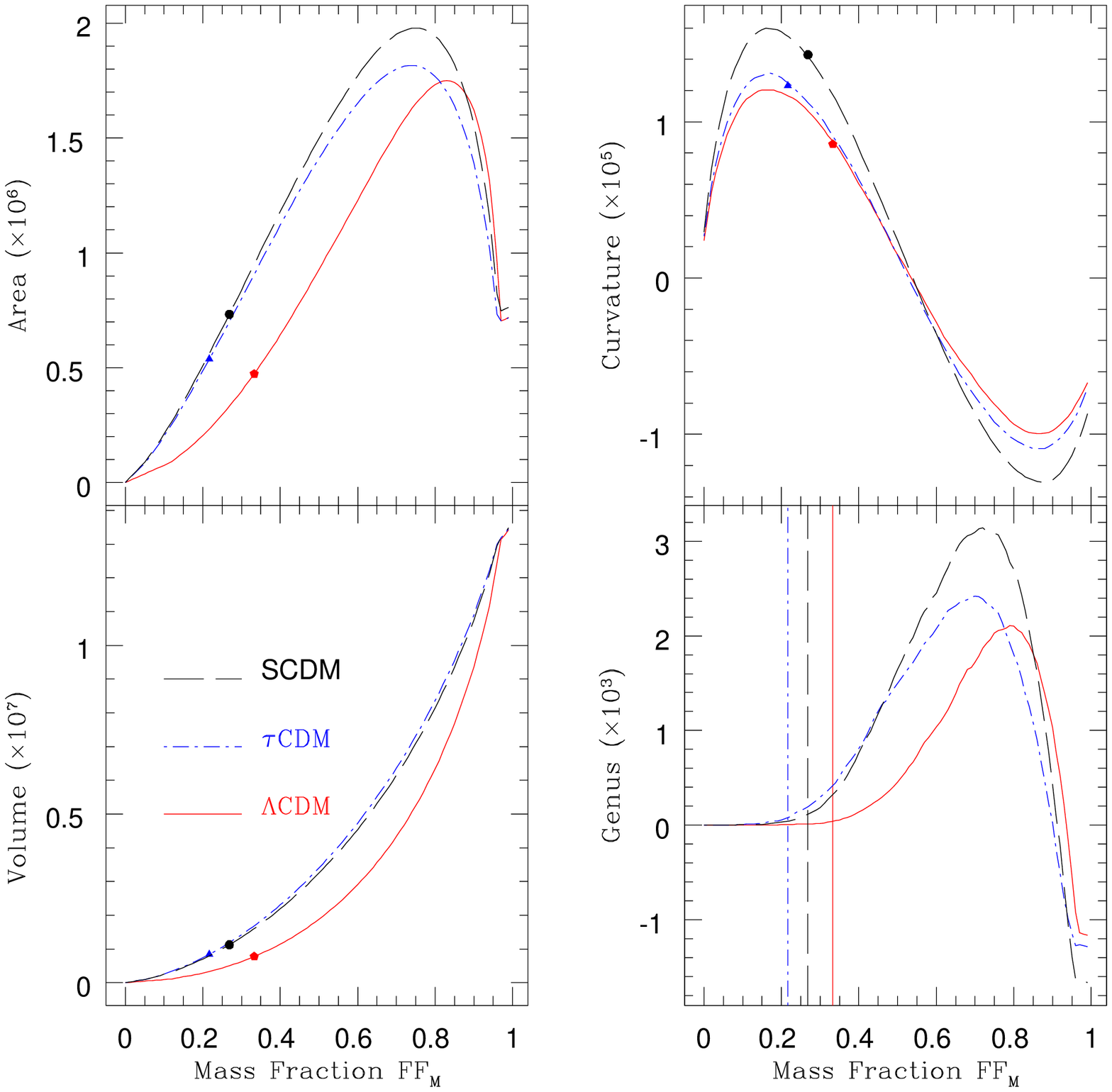}
  }
\caption{{Global Minkowski functionals for $\Lambda$CDM, $\T$CDM and
    SCDM are shown as functions of the mass fraction 
    defined as $FF_{\rm M} = M_{\rm total}^{-1}\int \rho ~(\geq
    {\rho_{\rm TH}}) ~dV$.  We see that some of the degeneracy 
    between the different models seen in Fig. \ref{fig:globalff} is
    broken when one plots the MFs in terms of the mass 
    fraction $FF_{\rm M}$ instead of the volume fraction $FF_{\rm V}$.
    The markers on the three curves show the
    mass fraction at the onset of percolation in each of
    the three cosmological models. (For the sake of greater clarity, vertical
    lines through the markers are shown in the lower right panel.)
    }}
\label{fig:globalmassff}
\end{figure*}
%%%%%%%%%%%%%%%%%%%
are shown in Figures \ref{fig:globalff}. This figure shows some
interesting features. The three cosmological models have appreciably
different morphology and hence can be distinguished from one another
on the basis of morphological measures.  Similarly to the percolation
statistic \citep{dsh92} the difference between models is brought out
much better if, instead of plotting the MFs against the volume
fraction (equivalently `volume filling factor'), we choose the `mass
fraction' defined as $FF_{\rm M} = M_{\rm total}^{-1}\int \rho ~(\geq
{\rho_{\rm TH}}) ~dV$.  (The `mass fraction' can be thought of as the
'volume or mass filling factor' \footnote{Since the density in the
  Lagrangian space is uniform there is no difference between the
  volume and mass fractions.}  in (initial) Lagrangian space, while
the `volume fraction' is the filling factor in (final) Eulerian
space.)  By employing this parameter for studying the global MFs, we
essentially probe the morphology of the iso-density contours (which
may refer to different thresholds of density or density contrast, but)
which enclose the same fraction of the total mass. Since clustering
tends to pack up the mass into progressively smaller regions of space,
such a study connects to aspects of gravitational clustering in a
direct way.  Thus, we see that at all the values of $FF_{\rm M}$ the
volume that encloses the same fraction of the total mass is least for
$\l$CDM and most for $\T$CDM. A further advantage of employing this
parameter is that the behaviour of the rest of the global MFs is no
longer restricted to follow the same pattern (like it is in case of
studying global MFs with the volume filling factor as the labelling
parameter). Thus, not only the peak, but also the peak-position and
the shape of the MF-curves are sensitive to the models being
investigated.  Thus, the mass fraction parametrisation could be more
useful in discriminating the models from one another as well as
comparing the models with the observations.
Before we commence our analysis of individual objects
(connected overdense/underdense regions)
we would like to draw the attention of the reader to
an issue in nomenclature. The `clusters' and `superclusters' referred
to further in this paper are elements constituting the `cosmic-web'
and defined on the basis of isodensity contours in N-body 
simulations. They should not be mistaken for the clusters and 
superclusters of galaxies seen in the sky.

\subsection{Cluster abundance and  percolation}
In this subsection we first discuss the percolation properties of
the three density fields being studied. In this context we study how
the total number of clusters and the fractional volume in the largest
cluster vary as we scan through a set of density levels corresponding
to equispaced fractions of volume and total mass.
%%%%%%%%%%%%%%%%%%%%
\begin{figure*}
\centering
\centerline{
  \includegraphics[width=5in]{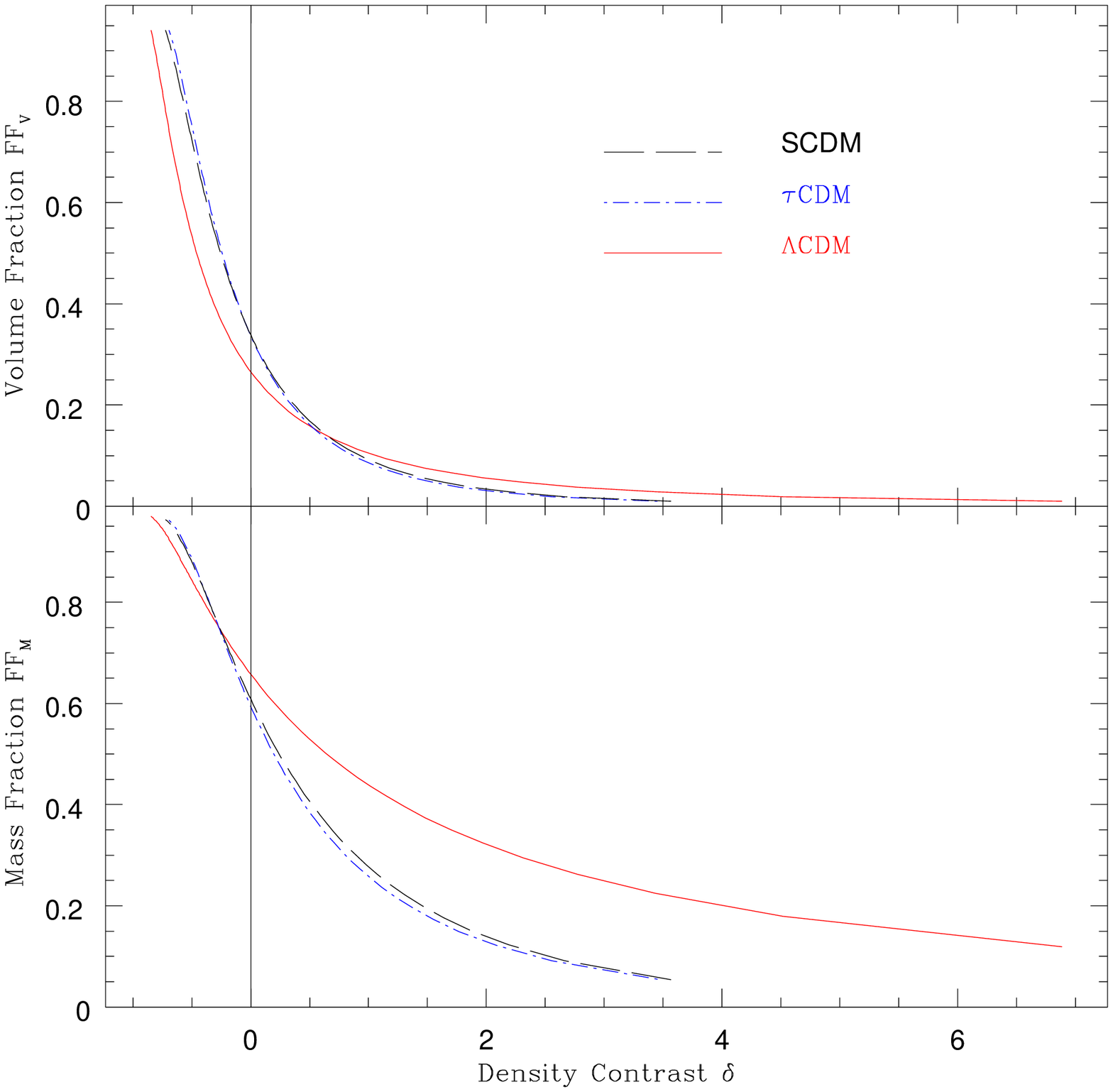}
  }
\caption{{Here we study the evolution of the volume fraction and the 
    mass fraction w.r.t. the density contrast for all the three
    models. $\Lambda$CDM shows maximum density contrast.  SCDM and
    $\tau$CDM form another class of models which have similar density
    contrasts but smaller than $\l$CDM. The vertical line across both
    the panels refers to the mean density, which can be used as a
    marker to study the mass fraction which is contained in the
    overdense volume (the volume occupied by regions above the mean
    density). We note that the overdense volume in $\L$CDM is
    $\sim$25\% which is about 10\% smaller than that occupied by the
    other two models.  At the same time, the mass that this volume
    encloses is $\sim$67\% and is about 5\% larger than that enclosed
    by the overdense volume in the other two models. }}
\label{fig:ffcontrast}
\end{figure*}
%%%%%%%%%%%%%%%%%%%%
Figure \ref{fig:ffcontrast} shows the volume fraction and mass
fraction as a function of the density contrast.  It is interesting to
note that SCDM and $\tau$CDM show the same pattern of behaviour which
differs from $\l$CDM. It is further to note that at high density
thresholds ($\delta \ggeq 1$) the filling factor (at a constant
density threshold) is greater in $\l$CDM than in SCDM or $\tau$CDM,
while exactly the reverse is true for underdense regions.  This
figure, which relates $FF_{\rm M,V}$ to $\delta$, serves as a reference point to
all of our subsequent percolation studies. Figure \ref{fig:ffcontrast}
also shows that, relative to other models, {\em more mass occupies
  less space in $\l$CDM}. Thus almost $67\%$ of the total mass in the
$\l$CDM universe resides in just $25 \%$ of the volume.  (In case of
$\tau$CDM and SCDM $\sim 62\%$ mass occupies $\sim 34\%$ volume.)
%%%%%%%%%%%%%%%%%%%%
\begin{figure*}
\centering
\centerline{
  \includegraphics[width=5in]{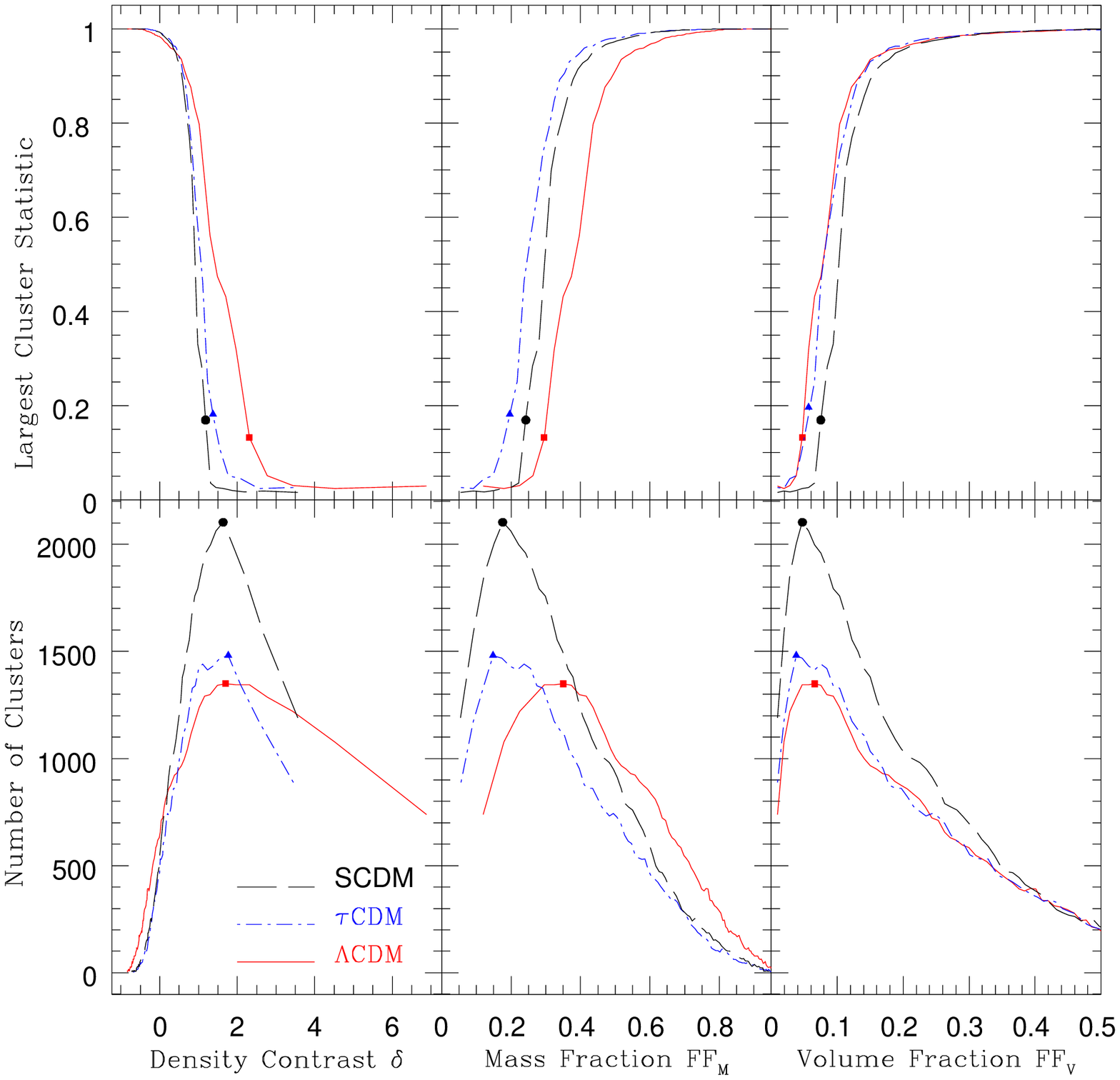}
  }
\caption{{Number of Clusters statistic (NCS) and the Largest Cluster 
    Statistic (LCS) are studied as functions of the density contrast
    $\delta$ (left panels), mass fraction $FF_{\rm M}$ (middle panels)
    and volume fraction, (equivalently volume filling factor),
    $FF_{\rm V}$ (right panels). The markers on the three curves, in
    the case of NCS, show values of $\delta_{\rm cluster-max}$ (left),
    $FF_{\rm M, max}$ (middle) and $FF_{\rm V, max}$ (right) at which
    the cluster abundance peaks. In the case of LCS, the markers
    indicate percolation values of $\delta_{\rm perc}$ (left),
    $FF_{\rm M, perc}$ (middle) and $FF_{\rm V, perc}$ (right) at
    which the largest cluster percolates by running through the entire
    simulation box.  }}
\label{fig:percolation}
\end{figure*}
%%%%%%%%%%%%%%%%%%%%

A statistical pair which help quantify the geometry of large scale
structure and its morphology are the `Number of Clusters Statistic'
(NCS) and the `Largest Cluster Statistic' (LCS); both are shown in
Figure \ref{fig:percolation}.  The Number of Clusters Statistic shows
cluster abundance as a function of the density contrast (lower left),
mass fraction (lower middle) and volume fraction (lower right).  The
Largest Cluster Statistic measures the fractional cluster volume
occupied by the largest cluster: 
\b 
LCS = \frac{V_{\rm LC}}{\sum_i V_i}, 
\label{eq:lcs_def}
\e 
where $V_i$ is the volume of the $i-$th cluster in the sample and
$V_{\rm LC}$ is the volume occupied by the largest cluster. Summation
is over all clusters evaluated at a given threshold of the density and
includes the largest cluster.  In Figure \ref{fig:percolation} LCS is
shown as a function of the density contrast (upper left), mass
fraction (upper middle) and volume fraction (upper right).

The square, triangle and circle in NCS denote critical values of the
parameters $\delta = \delta_{\rm cluster-max}$, $FF_{\rm V} = FF_{\rm
  cluster-max}$ and $FF_{\rm M}$ at which the cluster abundance peaks
in a give model. Similar etchings on the LCS curves denote values of
$\delta = \delta_{\rm perc}$, $FF_{\rm V} = FF_{\rm perc}$ and
$FF_{\rm M}$ at which the largest cluster first spans across the box
in at least one direction.  This is commonly referred to as the
`percolation threshold'.  It is important to note that the number of
clusters in SCDM is considerably greater than the number of clusters
in $\l$CDM at most density thresholds.  It is also interesting that
percolation takes place at higher values of the density contrast (and
correspondingly lower values of the volume filling factor) in the case
of $\l$CDM.  ($\delta_{\rm perc} \simeq 2.3$ for $\l$CDM, $\delta_{\rm
  perc} \simeq 1$ for SCDM and $\T$CDM.) Furthermore, while in the
case of $\l$CDM $\delta_{\rm cluster-max} \geq \delta_{\rm perc}$, in
the case of SCDM and $\T$CDM, $\delta_{\rm cluster-max} \lleq
\delta_{\rm perc}$. Thus in the latter two models, as the density
threshold is lowered, the cluster abundance initially peaks, then, as
the threshold is lowered further, neighboring clusters merge to form
the percolating supercluster. In $\l$CDM on the other hand, the
threshold at which the cluster abundance peaks also signals the
formation of the percolating supercluster. Having said this we would
like to add a word of caution: the analysis of one realization alone
does not allow us to asses reliably the statistical fluctuations in
the threshold of percolation. Therefore the final conclusion can be drawn
only after a study of many realizations of each model.

Clearly both the LCS threshold $\delta_{\rm perc}$ and the NCS
threshold $\delta_{\rm cluster-max}$ contain important information and
most of our subsequent description of supercluster morphology will be
carried out at one of these two thresholds.  Table \ref{tab:top10s123}
shows the Minkowski functionals and associated Shapefinders for the 10
largest (most voluminous) superclusters compiled at the percolation
threshold for the three cosmological models $\l$CDM, $\T$CDM and SCDM
while Table \ref{tab:top10s123} lists values for the Planarity and 
Filamentarity for these superclusters along with the mass that these
enclose and their genus value. 

Our $(239.5$h$^{-1}$Mpc$)^3$ $\l$CDM universe contains 1334 clusters
\& superclusters at the percolation threshold. 
%%%%%%%%%%%%%%%%%%%%%%
\begin{figure*}
\centering
\centerline{
\includegraphics[width=5in]{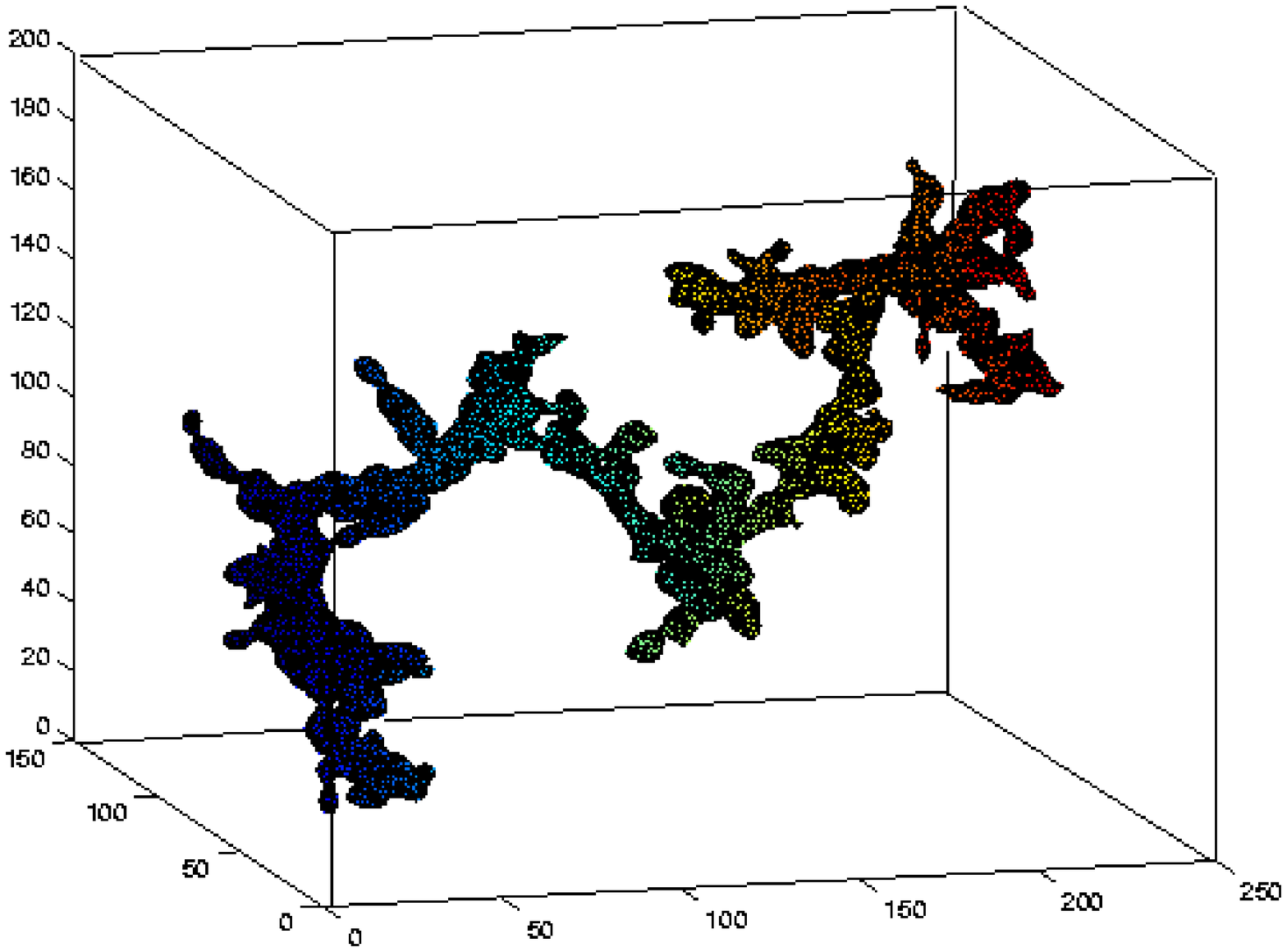}
}
\caption{{The largest (percolating) supercluster in $\l$CDM. 
    This cluster is selected at the density threshold which marks the
    onset of percolation ($\delta_{\rm perc} = 2.3$).  As demonstrated
    in the figure, the cluster at this threshold percolates through
    the entire length of the simulation box.  It is important to note
    that the percolating supercluster occupies only a small fraction
    of the total volume and its volume fraction (filling factor) is
    only $0.6\%$. Our percolating supercluster is a multiply
    connected and highly filamentary object. Its visual appearance
    is accurately reflected in the value of the Shapefinder diagnostic
assigned to this supercluster:
    ($T, B, L$) = (5.63, 7.30,
    70.03) $h^{-1}$Mpc and (P,F,G) = (0.13, 0.81, 6).
}}
\label{fig:perclcdm}
\end{figure*}
%%%%%%%%%%%%%%%%%%%%%%
Of these the smallest is quasi-spherical with a radius of a few
Megaparsec while the largest is extremely filamentary and percolates
through the entire simulation box (see figure \ref{fig:perclcdm}).
From Figures \ref{fig:perclcdm} \& \ref{fig:percolation} we find that
the percolating $\l$CDM supercluster is a slim but massive
object.\footnote{Surface visualization is a difficult task especially
  if we wish to follow tunnels through superclusters to check whether
  our visual impression of the genus agrees with that calculated using
  the Euler formula (\ref{eq:genus}). MATLAB has been used for surface
  plotting and the `reality' of tunnels is ascertained by rotating
  surfaces and by viewing the supercluster at various angles.}  It
contains $4.5\%$ of the total mass in the universe yet occupies only
$0.6\%$ of the total volume.  In SCDM the percolating supercluster
contains $ 4.4\%$ of the total mass and occupies $1.2\%$ of the total
volume, i.e., a similar mass occupies two times greater volume than
in the $\l$CDM.  The $\sim 10^3$ $\l$CDM clusters with $\delta \geq
\delta_{\rm perc}$ occupy $4.4\%$ of the total volume and contain
$33\%$ of the total mass in the universe.  Gravitational clustering
thus ensures that most of the mass in the $\l$CDM universe is
distributed in coherent filamentary regions which occupy very little
volume but contain much of the mass.  (Of the 1334 $\l$CDM clusters
and superclusters identified at the percolation threshold, the ten
most voluminous are extremely filamentary and contain close to $40\%$
of the total cluster mass (see table \ref{tab:top10s123}).  The
remaining $60\%$ of the mass is distributed amongst other 1324 objects
!)

The percolating supercluster is more filamentary and less planar in
$\l$CDM [$(P,F)_{\L CDM}$=(0.13,0.81)] than in SCDM
[$(P,F)_{SCDM}$=(0.14,0.78)]. The percolating supercluster of the
$\T$CDM is least filamentary and most planar with $(P,F)_{\T
  CDM}$=(0.15,0.70).  It is interesting that superclusters evaluated
at $\delta_{\rm perc}$ are more `porous' ({\em i.e.} have a larger
genus) in SCDM than their counterparts in $\l$CDM. For instance the
percolating supercluster in $\l$CDM has only 6 `holes' which tunnel
all the way through it, while 20 tunnels pass through the percolating
supercluster in SCDM; see Table \ref{tab:top10s123}.  (This number
increases as the density threshold is lowered as demonstrated in
Figure \ref{fig:shapespace-largecl}.)  Another interesting feature of
gravitational clustering appears to be that although the volume
occupied by clusters at $\delta_{\rm cluster-max}$ does not vary much
between models, the fractional mass contained in $\l$CDM clusters is
{\em almost twice} that in either $\T$CDM or SCDM; see Figure
\ref{fig:percolation}.

%%%%%%%%%%%%%%%%%%%%%%%%%%%%%%%%%%%% Table 4 %%%%%%%%%%%%%%%%%%%%%%%%%%%%%%%%%
\begin{table*}
\centering
\caption{The ten most voluminous superclusters (determined at the 
  percolaton threshold) are listed with the mass they enclose, their 
  associated Minkowski functionals
  (Volume, area, curvature and genus) and Shapefinders $T, B, L, P$ and $F$. 
  The first row in each cosmological model describes the percolating 
  supercluster and appears in boldface. It is interesting that in all three cosmological models the top
ten superclusters contain roughly $40\%$ of
the total mass in overdense regions with $\delta \geq \delta_{\rm perc}$.
(The precise numbers are $40\%$ for $\l$CDM, $37.8\%$ for SCDM
and $45.6\%$ in $\T$CDM.)
It should also be noted that the mass of a typical supercluster in $\L$CDM 
is somewhat smaller than that in the other two models mainly on account of the
fact that the adopted particle mass in
$\L$CDM simulations
is smaller than in simulations of SCDM and $\T$CDM. (The particle
mass in $\L$CDM is $6.86\times10^{10}h^{-1}M_{\odot}$ and that in SCDM and $\T$CDM is
$2.27\times10^{11}h^{-1}M_{\odot}$ \citep{jenkcdm}). It should be noted that the 
  interpretation of $L$ as the `linear length' of a supercluster can be 
  misleading for the case of superclusters having a large genus. In this 
  case $L\times (G+1)$ provides a more realistic  estimate of supercluster 
  length since it allows for its numerous twists and turns. The morphology
  of the objects is conveyed through their planarity $P$ and filamentarity 
  $F$. We note that the most voluminous and massive structures are highly
  filamentary in all the models. The largest supercluster is most filamentary
  in case of $\L$CDM and least so in case of $\T$CDM. }

\begin{center} 
\begin{tabular}{llllllllllllllr} \hline 
\multicolumn{1}{l} {Model} & 
\multicolumn{1}{l} {Mass} &
\multicolumn{1}{l} {Volume} & 
\multicolumn{1}{l} {Area} & 
\multicolumn{1}{l} {Curvature} & 
\multicolumn{1}{l} {Genus} & 
\multicolumn{5}{c} {Shapefinders} \\ 
 & (M/M$\odot$) & $(h^{-1}Mpc)^3$ & $(h^{-1}Mpc)^2$ & $h^{-1}$Mpc & & $T$ & $B$ & $L$ & $P$ & $F$ \\
 & & & & & & ($h^{-1}$Mpc) & ($h^{-1}$Mpc) & ($h^{-1}$Mpc) & & \\
\hline
$\Lambda$CDM & ${\bf 5.17\times10^{16}}$ & ${\bf 8.45\times10^4}$ &  ${\bf 4.5\times10^4}$ & ${\bf 6.16\times10^3}$ & {\bf ~~6} & {\bf 5.63} & {\bf 7.30} & {\bf 70.03} & {\bf 0.13} & {\bf 0.81} \\
$\delta=2.31$ & $1.87\times10^{16}$ & $3.21\times10^4$ & $1.7\times10^4$ & $2.33\times10^3$ & ~~~1 & 5.63 & 7.34 & 92.74 & 0.13 & 0.85 \\
      & $1.30\times10^{16}$ & $2.22\times10^4$ & $1.22\times10^4$ & $1.76\times10^3$ & ~~~1 & 5.47 & 6.93 & 70.00 & 0.12 & 0.82 \\
      & $1.41\times10^{16}$ & $2.13\times10^4$ & $1.02\times10^4$ & $1.29\times10^3$ & ~~~0 & 6.27 & 7.90 & 102.7 & 0.12 & 0.86 \\
      & $7.68\times10^{15}$ & $1.21\times10^4$ & $6.35\times10^3$ & $8.78\times10^2$ & ~~~0 & 5.71 & 7.22 & 69.95 & 0.12 & 0.81 \\
      & $6.75\times10^{15}$ & $1.19\times10^4$ & $6.82\times10^3$ & $1.02\times10^3$ & ~~~1 & 5.25 & 6.70 & 40.53 & 0.12 & 0.72 \\
      & $5.85\times10^{15}$ & $1.08\times10^4$ & $6.12\times10^3$ & $9.16\times10^2$ & ~~~0 & 5.3 & 6.68 &  72.90 & 0.11 & 0.83 \\
      & $7.13\times10^{15}$ & $1.01\times10^4$ & $5.46\times10^3$ & $8.19\times10^2$ & ~~~0 & 5.57 & 6.65 & 65.24 & 0.09 & 0.81 \\
      & $5.19\times10^{15}$ & $8.6\times10^3$  & $4.74\times10^3$ & $7.07\times10^2$ & ~~~0 & 5.44 & 7.00 &  56.33 & 0.10 & 0.79 \\
      & $5.04\times10^{15}$ & $8.1\times10^3$ & $4.25\times10^3$ & $5.71\times10^2$  & ~~~1 & 5.72 & 7.44 &  22.74 & 0.13 & 0.51 \\
\hline
$\tau$CDM & ${\bf 1.45\times10^{17}}$ & ${\bf 1.37\times10^5}$ & ${\bf 7.57\times10^4}$ & ${\bf 1.03\times10^4}$ & {\bf ~~19} & {\bf 5.42} & {\bf 7.36} & {\bf 40.94} & {\bf 0.15} & {\bf 0.70} \\
$\delta=1.37$ & $4.04\times10^{16}$ & $3.97\times10^4$ & $2.35\times10^4$ & $3.58\times10^3$ & ~~~2 & 5.07 & 6.57 & 94.93 & 0.13 & 0.87 \\ 
       & $3.16\times10^{16}$ & $2.98\times10^4$ & $1.72\times10^4$ & $2.48\times10^3$ & ~~~1 & 5.20 & 6.94 & 98.53 & 0.14 & 0.87 \\ 
       & $2.45\times10^{16}$ & $2.27\times10^4$ & $1.32\times10^4$ & $1.97\times10^3$ & ~~~1 & 5.16 & 6.71 &  78.47 & 0.13 & 0.84 \\
       & $1.99\times10^{16}$ & $1.77\times10^4$ & $9.38\times10^3$ & $1.29\times10^3$ & ~~~1 & 5.66 & 7.26 &  51.44 & 0.12 & 0.75 \\
       & $1.65\times10^{16}$ & $1.63\times10^4$ & $9.35\times10^3$ & $1.31\times10^3$ & ~~~2 & 5.24 & 7.14 & 34.73 & 0.15 & 0.66 \\
       & $1.74\times10^{16}$ & $1.39\times10^4$ & $6.69\times10^3$ & $8.14\times10^2$ & ~~~3 & 6.26 & 8.22 & 16.20 & 0.14 & 0.33 \\
       & $1.55\times10^{16}$ & $1.38\times10^4$  & $8.06\times10^3$ & $1.12\times10^3$ & ~~~1 & 5.13 & 7.21 & 44.45 & 0.17 & 0.72 \\
       & $1.46\times10^{16}$ & $1.34\times10^4$ & $7.8\times10^3$ & $1.16\times10^3 $ & ~~~2 & 5.15 & 6.72 & 30.77 & 0.13 & 0.64 \\
       & $1.48\times10^{16}$ & $1.32\times10^4$ & $7.32\times10^3$ & $1.04\times10^3$ & ~~~2 & 5.42 & 7.01 & 27.68 & 0.13 & 0.60 \\
\hline
SCDM & ${\bf 1.66\times10^{17}}$ & ${\bf 1.7\times10^5}$ & ${\bf 9.89\times10^4}$ & ${\bf 1.44\times10^4}$ & {\bf ~~20} & {\bf 5.14} & {\bf 6.88} & {\bf 54.57} & {\bf 0.14} & {\bf 0.78} \\
$\delta=1.19$ & $4.15\times10^{16}$ & $4.24\times10^4$ & $2.6\times10^4$ & $4.07\times10^3$ & ~~~2 & 4.90 & 6.38 & 107.94 & 0.13 & 0.89 \\
       & $3.43\times10^{16}$ & $3.52\times10^4$ & $2.04\times10^4$ & $2.93\times10^3$ & ~~~3 & 5.19 & 6.96 & 58.27 & 0.15 & 0.79 \\
       & $2.02\times10^{16}$ & $1.95\times10^4$ & $1.07\times10^4$ & $1.48\times10^3$ & ~~~3 & 5.47 & 7.21 & 29.47 & 0.14 & 0.61 \\
       & $1.85\times10^{16}$ & $1.94\times10^4$ & $1.12\times10^4$ & $1.56\times10^3$ & ~~~2 & 5.19 & 7.08 & 42.05 & 0.15 & 0.71 \\
       & $1.61\times10^{16}$ & $1.6\times10^4$ & $9.64\times10^3$ & $1.39\times10^3$ & ~~~3 & 4.97 & 6.95 & 27.58 & 0.17 & 0.60 \\
       & $1.34\times10^{16}$ & $1.36\times10^4$ & $7.59\times10^3$ & $1.08\times10^3$ & ~~~2 & 5.37 & 7.05 & 28.59 & 0.14 & 0.60 \\
       & $1.30\times10^{16}$ & $1.34\times10^4$ & $8.07\times10^3$ & $1.25\times10^3$ & ~~~2 & 4.97 & 6.47 & 33.07 & 0.13 & 0.67 \\
       & $1.29\times10^{16}$ & $1.25\times10^4$ & $7.11\times10^3$ & $9.98\times10^2$ & ~~~2 & 5.27 & 7.12 & 26.50 & 0.15 & 0.58 \\
       & $1.30\times10^{16}$ & $1.21\times10^4$ & $6.87\times10^3$ & $9.88\times10^2$ & ~~~2 & 5.27 & 6.95 & 26.22 & 0.14 & 0.58 \\
\hline
\label{tab:top10s123}
\end{tabular}
\end{center}
\end{table*}

\subsection{Cluster Morphology}
Cosmic density fields contain a wealth of information.  As
demonstrated in Fig. \ref{fig:percolation}, at density thresholds
significantly lower than $\delta_{\rm perc}$ most clusters merge to
form a single percolating supercluster.  On the other hand, the
slightly larger NCS density contrast, $\delta_{\rm cluster-max}$,
provides an excellent threshold at which to study the morphology of
individual objects since it is precisely at $\delta_{\rm cluster-max}$
that the cluster abundance peaks.  We study the morphology and
topology of large scale structure in a two-fold manner: (i) properties
of all superclusters are analysed at one of the two thresholds:
$\delta_{\rm perc}, \delta_{\rm cluster-max}$; (ii) the morphology of
the largest supercluster is extensively probed as a function of the
density contrast.

%%%%%%%%%%%%%%%%%%%
\begin{figure*}
\centering
\centerline{
  \includegraphics[width=5in]{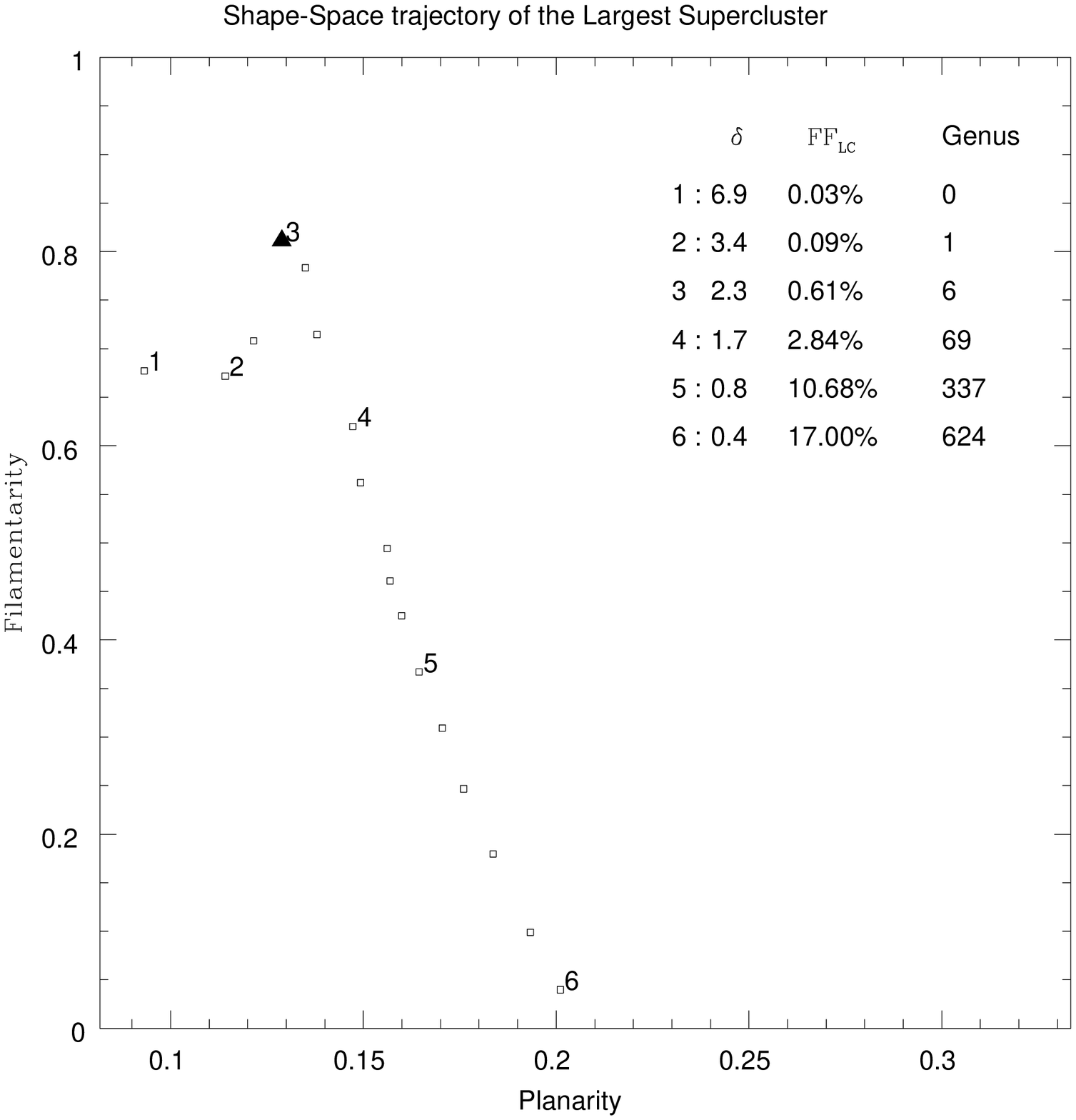}
}
\caption{{The morphological evolution of the
    largest supercluster in $\L$CDM is shown as a series of open
    squares in shape-space $\lbrace F,P \rbrace$.  Each square
    corresponds to a different value of the density threshold which is
    progressively lowered from a large initial value ($\delta \simeq
    6.9$; left most square) until the mean density level ($\delta =
    0$; lower most square). The legend lists the density contrast, the
    associated volume fraction and the genus of the largest
    supercluster at six monotonically decreasing values of the density
    contrast (1 $\to$ 6). At the highest ($\delta \simeq 6.9$), the
    largest cluster appears to have a large filamentarity and a small
    planarity.  The solid triangle (labelled 3) refers to the
    percolation threshold at which the largest cluster first spans
    across the simulation box.  After percolation, the filamentarity
    rapidly decreases from a maximum value of 0.81 to 0.0 as the
    threshold is lowered from $\delta \simeq 2$ to $\delta \simeq 0$.
    The decline in filamentarity of the largest cluster is accompanied
    by growth in its complexity as revealed by its genus value, e.g., the
    fractional volume occupied by the largest cluster at the mean
    density is $\sim$26\% and its genus exceeds a thousand.  }}
\label{fig:shapespace-largecl} 
\end{figure*}
%%%%%%%%%%%%%%%%%%
Figure \ref{fig:shapespace-largecl} shows the evolution of the
morphology and topology of the largest cluster as we scan through
different threshold levels of the $\l$CDM density field.  The density
contrast, volume fraction and genus for a few of these levels have
been labelled and are given in the right corner of the figure.  We
note that at high density thresholds the largest cluster occupies a
very small fraction of the total volume and its shape is characterized
by significant filamentarity ($\sim$0.67) and negligible planarity
($\sim$ 0). 
% The phase with large values of the filamentarity is spread
% over a small plateau which abruptly ends when its volume fraction
% reaches $\sim$5\%. The plateau region includes the percolation
% threshold $\delta_{\rm perc}$ (shown as a solid triangle). 
We note a sharp increase in the filamentarity of the largest cluster
as we approach the percolation threshold (shown in the figure as a
solid triangle). At smaller values of the density contrast, $\delta
\lleq 3$, supercluster filamentarity rapidly drops while its planarity
considerably increases.  The drop in filamentarity of the supercluster
is accompanied by a growth in its complexity with the result that, at
moderately low thresholds ($\delta \sim 0.5$), the percolating
supercluster can contain several hundred tunnels. At very low
thresholds ($ \delta \lleq 0.5$) the percolating supercluster is an
isotropic object possessing negligible values of both planarity (P)
and filamentarity (F).

Since $\delta_{\rm perc}$ and $\delta_{\rm cluster-max}$ contain
information pertaining to morphology and connectivity, we compile
partial MFs for individual clusters at both these thresholds for the
three cosmological models $\l$CDM, SCDM and $\T$CDM.  We find it
convenient to work with the cumulative probability function (CPF)
which we define as the normalized count of clusters with the value of
a quantity Q to be greater than q at a given value 'q'.  The quantity
Q could be one of the MFs or one of the Shapefinders. We shall study
the dependence of CPF with Q on a log-log scale.

%%%%%%%%%%%%%%%%%%%%
\begin{figure*}
\centering
\centerline{
  \includegraphics[width=5in]{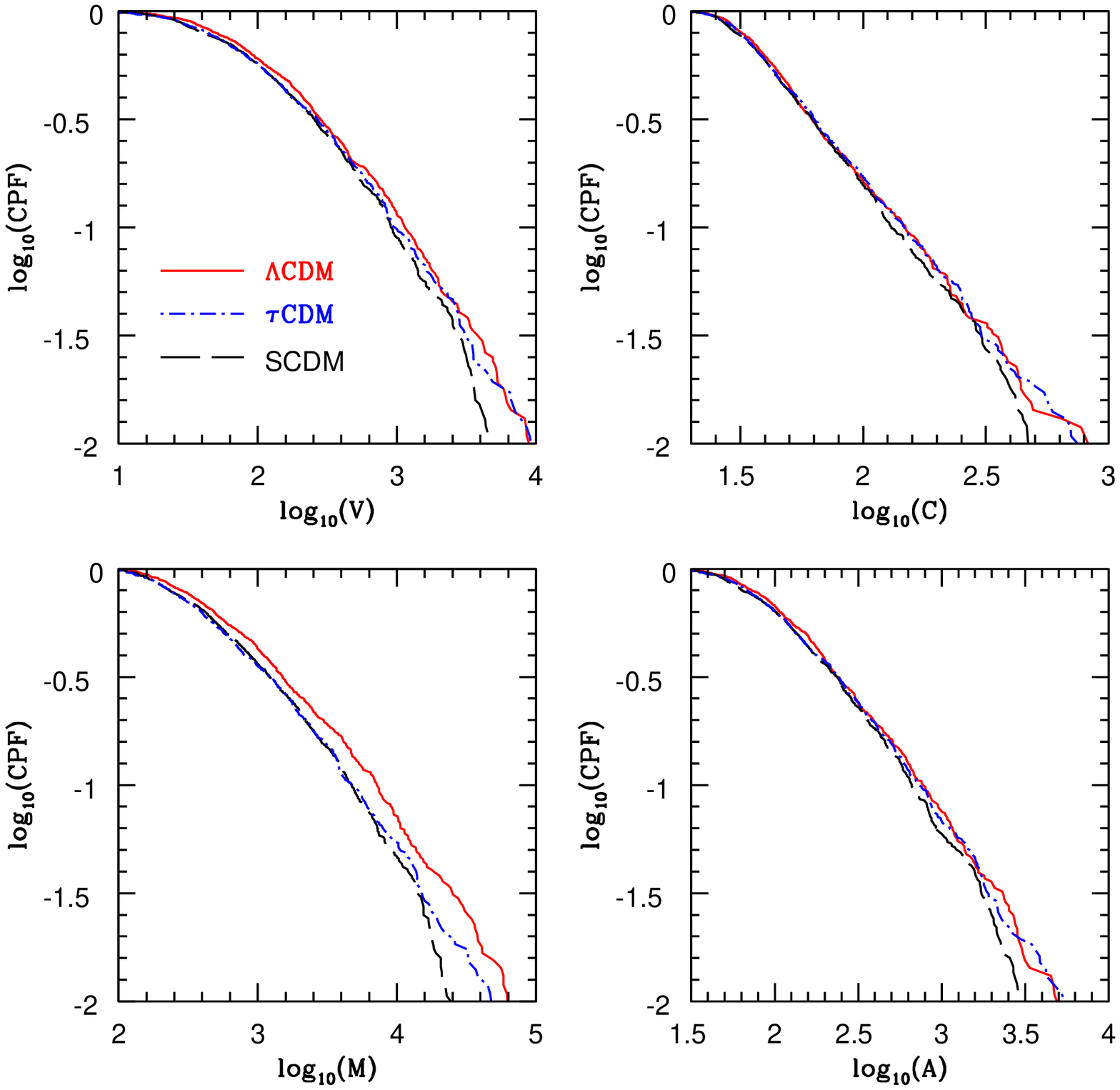}
  }
\caption{{Cumulative Probability Function (CPF) is shown as a function
    of total mass in superclusters (lower left) and as a function of
    the three Minkowski functionals volume (V), surface area (A) and
    mean extrinsic curvature (C).  Superclusters are defined to be
    connected overdense regions lying above the NCS threshold
    $\delta_{\rm cluster-max}$.  (The cluster abundance peaks at
    $\delta = \delta_{\rm cluster-max}$ (see figure
    \ref{fig:percolation}) which makes this a convenient threshold at
    which to probe morphology.) The colour type and line type hold
    the same meaning across all the panels.}}
\label{fig:cpfmfncs}
\end{figure*}
%%%%%%%%%%%%%%%%%%%%
%%%%%%%%%%%%%%%%%%%%%
\begin{figure*}
  \centering
  \centerline{
    \includegraphics[width=5in]{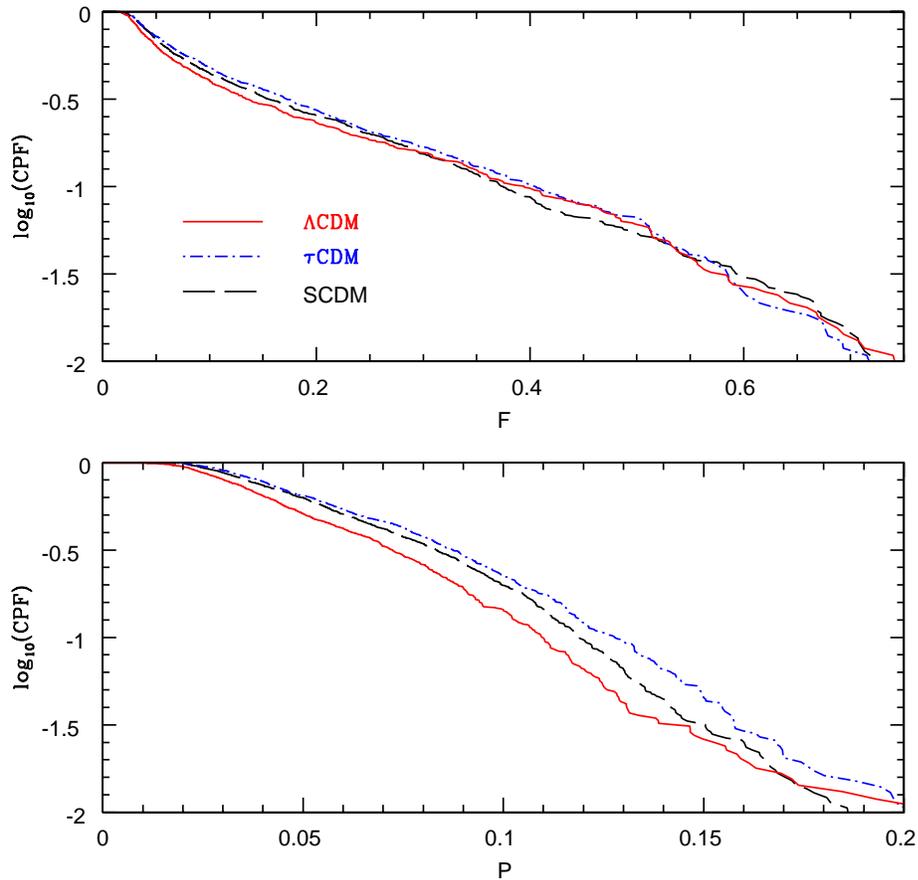}
    }
  \caption{{Cumulative Probability Function (CPF) for 
      filamentarity ($F$) and the planarity ($P$) of superclusters
      selected at the density threshold $\delta_{\rm cluster-max}$.
      The colour type and line type hold the same meaning for both the
      panels.}}
  \label{fig:cpffpncs}
\end{figure*}
%%%%%%%%%%%%%%%%%%%%
\begin{table*}
\centering
\caption{ The Kolmogorov-Smirnov statistic $d$ (first row for each pair
of models) and the probability (second row)
that the two data sets are drawn from the same distribution.
The models $\l$CDM, $\T$CDM and SCDM are compared on the basis of
three Minkowski functionals: volume (V),
area (A), integrated mean curvature (C), as well as 
mass (M) and the Shapefinders:
thickness (T), breadth (B),
length (L),  planarity (P), and filamentarity (F).
Boldface highlights the three {\em smallest probabilities}
for each pair of models.}
\begin{center} 
\begin{tabular}{l|llll|llll|ll} \hline
Models & M & V & A & C & $T$ & $B$ & $L$& P & F \\
\hline
$\Lambda$CDM - $\tau$CDM &
0.093 & 0.067 & 0.059 & 0.037 & 0.082 & 0.070 & 0.037 & 0.15 & 0.101\\
& ${\bf  9.4\times10^{-4}}$  &  0.036  &  $0.094$  &  $0.56$ &
$5.0\times10^{-3}$  &  0.026  &  $0.56$ &
${\bf 6.5\times10^{-9}}$  &  ${\bf 2.2\times10^{-4}}$\\
\hline
$\Lambda$CDM - SCDM & 
0.077 & 0.060 & 0.057 & 0.049 & 0.079 & 0.067 & 0.049 & 0.12& 0.074\\
& ${\bf4.6\times10^{-3}}$ &  0.048  &  0.069  &  0.17 &
${\bf 2.9\times10^{-3}}$  &  0.018  &  0.16  &
${\bf 3.2\times10^{-7}}$  & $6.3\times10^{-3}$\\
\hline
$\tau$CDM - SCDM & 
0.034 & 0.024 & 0.032 & 0.041 & 0.021 & 0.023 & 0.042 & 0.052& 0.046\\
& 0.54  &  0.90  &  0.61  &  0.32  &
  0.97  &  0.92  &  {\bf  0.29} &
{\bf 0.096}  & {\bf 0.19} \\
\hline
\label{tab:ks}
\end{tabular}
\end{center}
\end{table*}
%%%%%%%%%%%%%%%%%%%%
\begin{figure*}
  \centering
  \centerline{
    \includegraphics[scale=0.9,trim=20 140 5 140,clip]{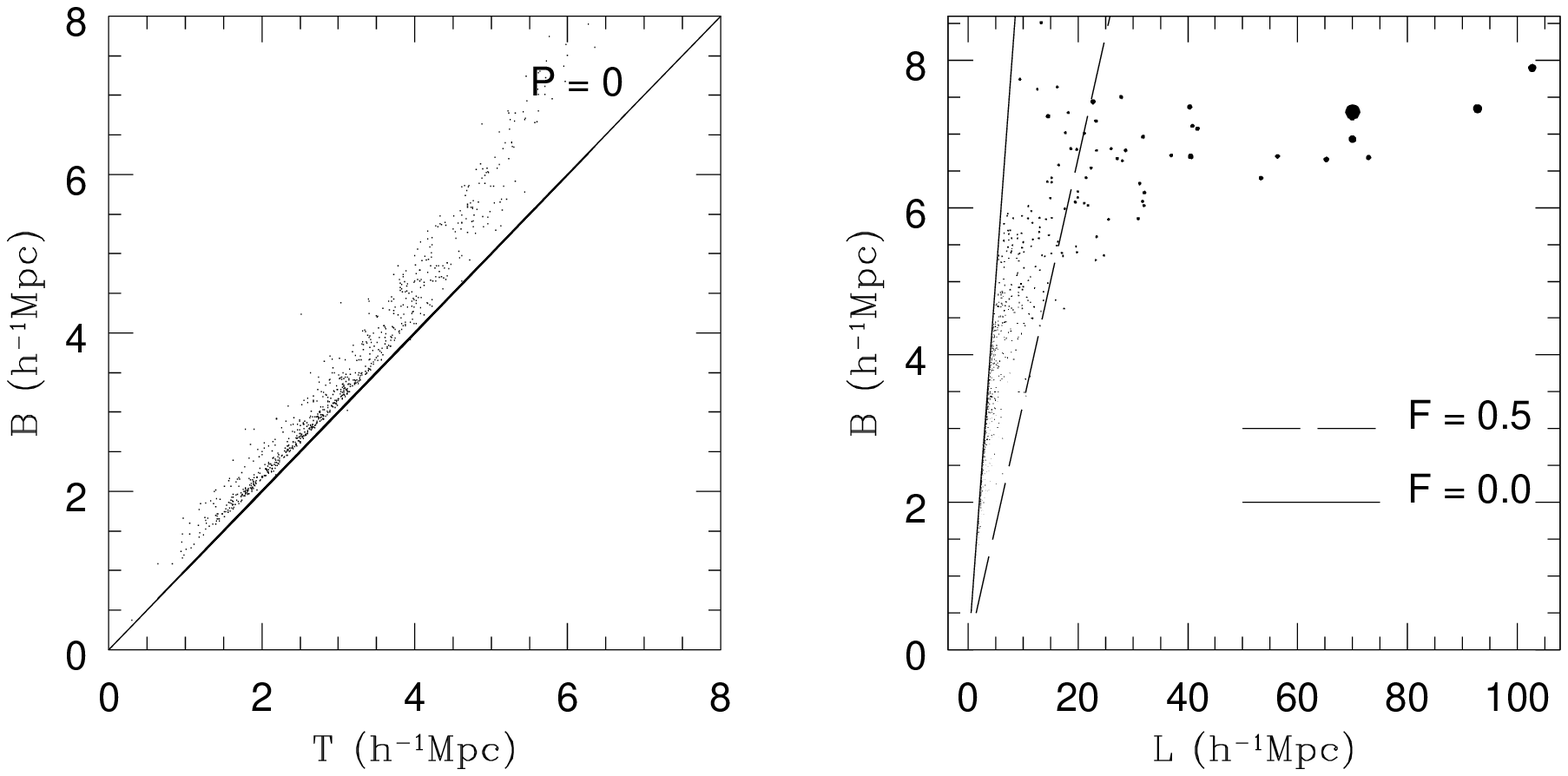}
    }
\caption{{Scatter plot for the
    pair of Shapefinders $T$, $B$ (left panel) and $B$, $L$ (right
    panel) defining the morphology of clusters/superclusters in the
    $\l$CDM model.  The strong correlation between $T$ and $B$ in the
    left panel near the line P = 0 indicates that two of the three
    dimensions defining a cluster are equal and of the same order as
    the correlation length ($\simeq$ few Mpc.).  Judging from the left
    panel we find that clusters/superclusters in $\l$CDM are either
    quasi-spherical or filamentary (since both satisfy $T \simeq B
    \Rightarrow$ P $\simeq 0$).  The degeneracy between spheres and
    filaments is lifted by the right panel which is a mass-weighted
    scatter plot for the Shapefinders $B,L$. Each dot in this panel
    refers to a cluster and its area is proportional to the fraction
    of mass in that cluster.  The concentration of points near the
    line F = 0 ($B = L$) reflects the fact that a large number of
    smaller clusters are quasi-spherical.  The more massive
    structures, on the other hand, tend to be filamentary and the
    largest and most massive supercluster has $F = 0.81$.  All objects
    are determined at the percolation threshold.}}
\label{fig:corr_5fs123}
\end{figure*}
%%%%%%%%%%%%%%%%%%%%%%
Our results are presented in Figures \ref{fig:cpfmfncs}, and
\ref{fig:cpffpncs} for clusters compiled at the NCS threshold
$\delta_{\rm cluster-max}$. For large values of the Minkowski
functionals ($V,A,C$) the CPF declines rapidly.  Although the curves
may look similar, some of them are statistically very different from
their peers. The first four columns of Table \ref{tab:ks} show the
results of the Kolmogorov-Smirnov test applied to the distribution of
the MFs.  (We do not carry out a similar exercise for the genus since
the vast majority of clusters at the NCS threshold are simply
connected.) In particular the CPF of the masses clearly distinguish
the $\Lambda$CDM model from SCDM \& $\T$CDM, in agreement with Fig.
\ref{fig:cpfmfncs}.

Figure \ref{fig:cpffpncs} shows the Cumulative Probability Function
for planarity and filamentarity in our cluster sample. The last two
columns of Table \ref{tab:ks} show the results of the KS test for
these statistics. For all three pairs they are among three best
discriminators of the models. The value of KS statistic $d$ suggests
that the signal comes from relatively small values of P ($< 0.1$) and
F ($< 0.25$). The more conspicuous differences in the tail of the
distribution is not statistically significant due to poor statistics.
Clusters in $\Lambda$CDM are the least anisotropic, a fact which
corresponds to their relatively early formation and therefore longer
evolution.  From Figure \ref{fig:cpffpncs} we also find that clusters
at the NCS threshold are significantly more filamentary than they are
planar.  This appears to be a generic prediction of gravitational
clustering as demonstrated by \cite{ashz82,ksh83,sss96} and
\cite{bondo}.

Figure \ref{fig:corr_5fs123} is a scatter plot of Shapefinders $T,B,L$
for clusters in $\l$CDM defined at the percolation threshold.  The
strong correlation between $T$ and $B$ in the left panel indicates
that two (of three) dimensions defining any given cluster assume
similar values and are of the same order as the correlation length.
(Note that $T \simeq B \simeq 5$h$^{-1}$ Mpc for the largest
superclusters in Table \ref{tab:top10s123}.)  The clustering of
objects near $T \simeq B$ ($P \simeq 0$) in this panel suggests that
our clusters/superclusters are either quasi-spherical or filamentary.
The scatter plot between $B$ \& $L$ in the right panel of Figure
\ref{fig:corr_5fs123} breaks the degeneracy between spheres and
filaments. The mass-dependence of morphology is highlighted in this
panel in which larger dots denote more massive objects. This figure
clearly reveals that more massive clusters/superclusters are, as a
rule, also more filamentary, while smaller, less massive objects, are
more nearly spherical. The concentration of points near the `edge' of
the scatter plot in the right panel, arises due to an abundance of low
mass compact quasi-spherical objects with $L \simeq B$ ($F \simeq 0$).
The large number of massive superclusters with $F > 0.5$ serves to
highlight the important fact that the larger and more massive elements
of the supercluster chain consist of highly elongated filaments as
much as $\sim 100$h$^{-1}$ Mpc in length, with mean diameter $\sim
5$h$^{-1}$ Mpc (see Table \ref{tab:top10s123}).  It is important to
reiterate that almost 40\% of the total overdense mass resides in the
ten largest objects listed in table \ref{tab:top10s123}, while the
remaining 60\% is distributed among 1324 clusters.

We also find short filaments to be generally thinner than
longer ones in agreement with predictions made by the adhesion model
with regard to the formation of hierarchical filamentary structure
during gravitational clustering \citep{kpsm92}.

\begin{figure*}
\centering
\centerline{
  \includegraphics[scale=0.9,angle=0]{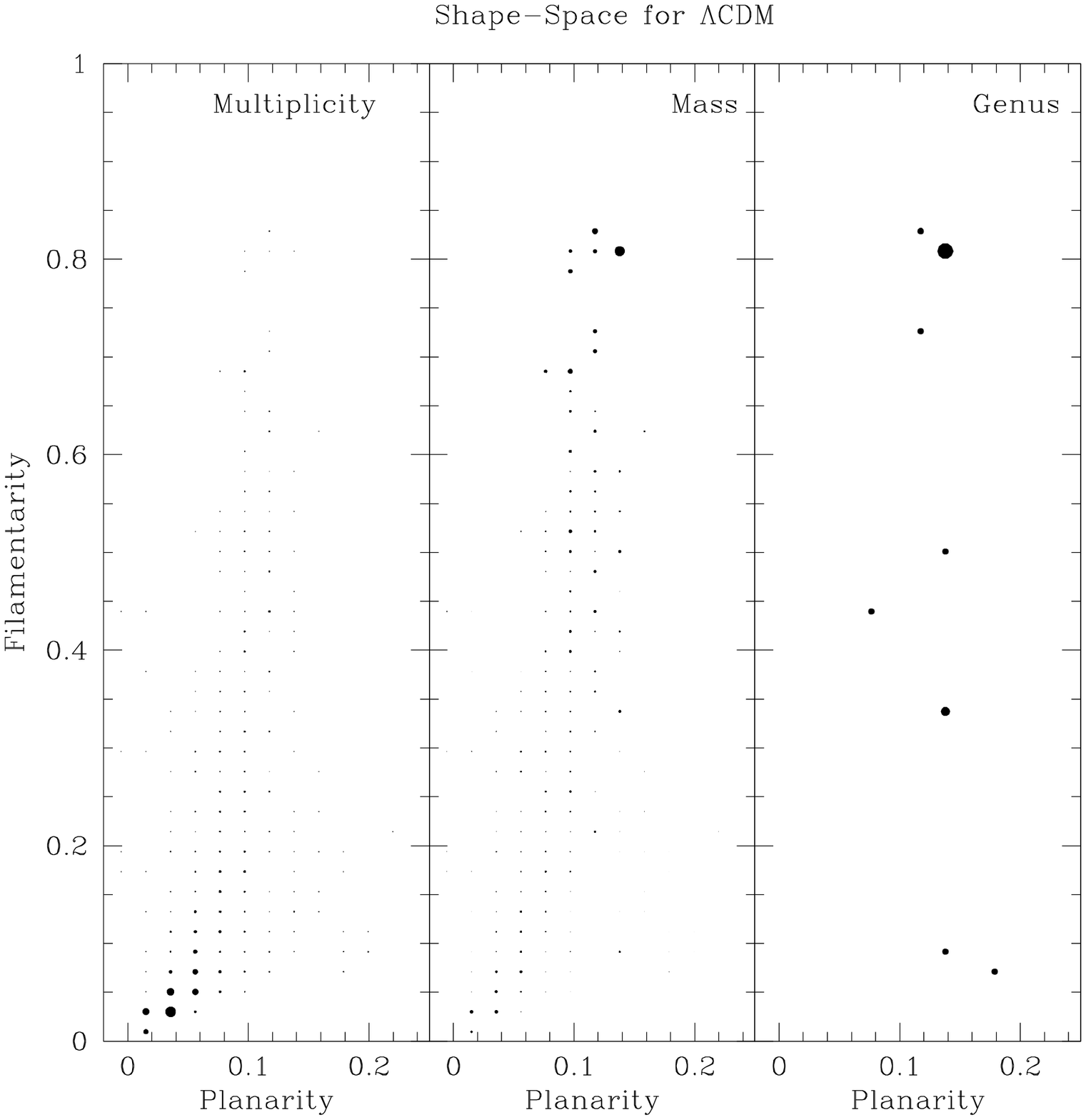}
  }
\caption{{Shape-Space for $\l$CDM. Left panel
    (multiplicity): The dots in this panel have area proportional to
    the number of clusters with a given (binned) value of
    Filamentarity and Planarity.  Center panel (mass): The dots have
    area proportional to the total mass contained in clusters having a
    given (binned) value of Filamentarity and Planarity. Note that
    more massive superclusters are also more filamentary.  Right panel
    (Genus): The dots have area proportional to the total genus value
    of superclusters with a given (binned) value of Filamentarity and
    Planarity.  This figure demonstrates the correlation between the
    mass of a supercluster, its shape and its genus. More massive
    superclusters are, as a rule, very filamentary and also
    topologically multiply connected.  As in the previous figure, all
    objects are determined at the percolation threshold.}}
\label{fig:shape-space}
\end{figure*}

To further probe the morphology of clusters and superclusters we
define the notion of shape-space in Figure \ref{fig:shape-space}.
Shape-space is two dimensional with the planarity (of a cluster)
plotted along the x-axis and its filamentarity along the y-axis.  (One
can also incorporate a third dimension showing the genus.)  The first
panel in Figure \ref{fig:shape-space} is a scatter plot of $P$ and $F$
for clusters in $\l$CDM. The sizes of dots in the middle panel are
proportional to cluster mass. We note that the most massive structures
are also very filamentary. In the right panel we try to relate the
shape of the structures with their topology by scaling the size of the
dots with the genus value of clusters having a given morphology
($P,F$). As shown here, clusters which are multiply connected
(larger dots are indicative of more complicated topologies) are also more
filamentary.  Together, the three panels show us that
more massive superclusters are frequently very filamentary and often
also topologically quite complex. We also see that a large number of
less massive superclusters are simply connected and
prolate. These structures are a few Mpc across along their
two shorter dimensions
and $\sim 20$ Mpc along the third, and therefore have appreciable filamentarity
($F \sim 0.3$). 
%%%%%%%%%%%%%%%%%%%%
%%%%%%%%%%%%%%%%%%%%%%%%%%%%%%

\section{Discussion and Conclusions}

This paper presents a new technique for studying the geometrical and
topological properties of large scale structure using the Minkowski
functionals (MFs). Given a density field reconstructed by
appropriately smoothing a point data set consisting (say) of a
distribution of galaxies, our ansatz: (i) constructs closed polyhedral
surfaces of constant density corresponding to excursion sets of a
density field, using for this purpose a surface generating
triangulation scheme (SURFGEN); (ii) SURFGEN evaluates the Minkowski
functionals (volume, surface area, extrinsic curvature and genus) thus
providing full morphological and topological information of three
dimensional iso-density contour surfaces corresponding to a given data
set.  Evaluated in this manner, the Minkowski functionals can be used
to study the properties of individual objects lying either above
(clusters, superclusters) or below (voids) a given density threshold.
They can also be used to study the morphological properties of the
full supercluster-void network at (say) the percolation threshold. The
ratio's of Minkowski functionals (Shapefinders) are used to probe the
shape of iso-density surfaces which sample the distribution of large
scale structure at different thresholds of the density. (The highest
density thresholds correspond to galaxies and clusters of galaxies,
moderate thresholds correspond to superclusters while the lowest
density thresholds characterize voids.)  The performance of our ansatz
has been tested against both simply and multiply connected surfaces
such as the sphere, triaxial ellipsoid and triaxial torus.  These
three eikonal bodies can be smoothly deformed to give surfaces which
are spheroidal, pancake-like (oblate) and filament-like (prolate) etc.
Analytically known values of the Minkowski functionals for these
surfaces allow us to test both our surface modelling scheme and our
evaluation of Minkowski functionals from triangulation.  Remarkably,
we find that volume, area and integrated extrinsic curvature are
determined to an accuracy of better than $1\%$ on all length-scales
for both simply and multiply connected surfaces.  The integrated
intrinsic curvature (genus) is evaluated exactly. We also find that
SURFGEN is remarkably accurate at calculating the MFs for Gaussian
random fields (see the appendix). Having validated the performance of
the code against eikonal surfaces and Gaussian random fields, we apply
our method to cosmological simulations of large scale structure
performed by the Virgo consortium.  We study the geometry and topology
of large scale structure in three cosmological scenarios - $\l$CDM,
$\T$CDM and SCDM.  All three cosmologies are analysed at the present
epoch (z=0) using global MFs, partial MFs and Shapefinders. (The
redshift evolution of geometry and topology will be discussed in a
companion paper.)  Our main conclusions are summarized below:

\begin{itemize}
\item Using the Minkowski functionals we show that, like other
  diagnostics of clustering, supercluster morphology too is sensitive
  to the underlying cosmological parameter set characterizing our
  universe.  Although the three cosmological models considered by us,
  $\l$CDM, $\T$CDM and SCDM display features which are qualitatively
  similar, the geometrical and morphological properties of these
  models are sufficiently distinct to permit differentiation using
  MFs.  We demonstrate that studying Minkowski Functionals using
  mass-parameterization significantly enhances their discriminatory
  power.

\item SURFGEN employs a new contour-based method for determining
  topology.  This method evaluates the genus for an iso-density contour
  surface directly, by applying the Euler formula Eq.(\ref{eq:genus}) to
  the triangulated surface, and presents a significant improvement over
  earlier grid-based methods of determining topology
  \citep{melotopo90}.
  
\item A study of percolation and cluster abundance reveals several
  interesting aspects of the gravitational clustering process.  For
  all models, on decreasing the density threshold to progressively
  smaller values one reaches the critical percolation threshold at
  which the largest supercluster runs through (percolates) the
  simulation box. Percolation is reached at moderate values of the
  density contrast ranging from $\delta_{\rm perc} \simeq 2.3$ for
  $\l$CDM to $\delta_{\rm perc} \simeq 1.2$ for SCDM. The abundance of
  clusters reaches a maximum value at (or very near) the percolation
  threshold and the percolating supercluster occupies a rather small
  amount of space in all three cosmological models. Thus the fraction
  of total simulation-box volume contained in the percolating
  supercluster is least in $\l$CDM ($0.6\%$) and greatest in SCDM
  ($1.2\%$).  When taken together, all overdense objects at the
  percolation threshold occupy $4.4\%$ of the total volume in the
  $\l$CDM model.  For comparison, the volume fraction in overdense
  regions at the percolation threshold is $\sim 16\%$ in an idealised,
  continuous Gaussian random field \citep{shandzed89}.  This fraction
  can increase upto $\sim 30\%$ for Gaussian fields generated on a
  grid \citep{ys96,sss97}.  The fact that clusters \& supercluster
  occupy a very small fraction of the total volume appears to be a
  hallmark of the gravitational clustering process which succeeds in
  placing a large amount of mass ($\sim 30\%$ of the total, in the
  case of $\l$CDM) in a small region of space ($\sim 4\%$).  The low
  filling fraction of the percolating supercluster in $\l$CDM
  ($0.006$) strongly suggests that this object is either planar or
  filamentary \citep{sss97} and a definitive answer to this issue is
  provided by the Shapefinder statistic.
  
\item Shapefinders were introduced to quantify the visual impression
  one has of the supercluster-void network of being a cosmic web of
  filaments interspersed with large voids \citep{sss98}.  By applying
  the Shapefinder statistics via SURFGEN to realistic N-body
  simulations we have demonstrated that: (i) Most of the mass in the
  universe is contained in large superclusters which are also
  extremely filamentary; the vast abundance of smaller clusters and
  superclusters tends to be prolate or quasi-spherical.  (ii) Of the
  three cosmological models the percolating supercluster in $\l$CDM is
  the most filamentary ($F \simeq 0.81$) and the supercluster in
  $\T$CDM the least ($F \simeq 0.7$). (iii) The percolating
  supercluster in $\l$CDM is topologically a much simpler object than
  its counterpart in $\T$CDM, with the former having only $\simeq 6$
  tunnels compared to $\simeq 19$ in the latter. Other differences
  between the models are quantified in Figures \ref{fig:globalff} $-$
  \ref{fig:shape-space}.  We also show that among various
  morphological parameters, the Planarity and Filamentarity of
  clusters and superclusters is one of the most powerful statistics to
  discriminate between the models (see Table \ref{tab:ks}).

\end{itemize}
To summarize, this paper has demonstrated that Minkowski functionals
and Shapefinders evaluated using SURFGEN provide sensitive probes of
the geometry, topology and shape of large scale structure and help in
distinguishing between rival cosmological models. Having established
the strength and versatility of SURFGEN it would clearly be
interesting to apply it to other important issues, related to those
addressed in this paper, including: (i) geometry and morphology of
underdense regions (voids); (ii) geometry and morphology of strongly
overdense regions (clusters); (iii) redshift-space distortions of
supercluster/void morphology; (iv) the time-evolution of the
supercluster-void network; (v) analysis of superclusters and voids in
fully three dimensional surveys such as 2dFGRS and SDSS; (vi) SURFGEN
is also likely to provide useful insights in other physical and
astrophysical situations in which matter is distributed
anisotropically such as in the interstellar medium
(\citet{filaments1,filaments}).  We hope to return to some of these
subjects in the near future.

\section{Acknowledgements}

We acknowledge useful discussions with Andrew Benson, Shaun Cole,
Suman Datta, Carlos Frenk, Adrian Jenkins, Madan Rao, C.D.Ravikumar,
Tarun Deep Saini, Yuri Shtanov, Volker Springel and David Weinberg.
Our insights into programming techniques of computational geometry
were honed by the excellent monograph on the subject by Joseph
O'Rourke.  Part of this work was done during a visit of JVS, VS and SS to
the Department of Physics $\&$ Astronomy, Cardiff University, Cardiff
under a PPARC Grant Ref. PPA/V/S/1998/00041. JVS is supported by the
research fellowship of the Council of Scientific and Industrial
Research (CSIR), India. VS and SS acknowledge support from the Indo-US
collaborative project DST/NSF/US (NSF-RP087)/2001.  The simulations
studied in this paper were carried out by the Virgo Supercomputing
Consortium using computers based at Computing Centre of the Max-Planck
Society in Garching and at the Edinburgh Parallel Computing Centre.
The data are publicly available at www.mpa-garching.mpg.de/NumCos.

\appendix

\section{Gaussian Random Fields and the Role of Boundary Conditions}
This appendix demonstrates the great accuracy with which SURFGEN
determines Minkowski functionals for Gaussian random fields (hereafter
GRFs).

Before embarking on our discussion, we briefly describe the samples
that we use and summarise the analytical results against which the
performance of SURFGEN is to be tested.

We work with three realizations of a gaussian random field with a
power law power spectrum (n = $-$1) on a $128^3$ grid. Each
realization of the field is smoothed with a gaussian kernel of length
$\lambda =$ 2.5 grid-units. All the fields are normalised by the
standard deviation.  This leaves them with zero mean and unit
variance.  The MFs are evaluated at a set of equispaced levels of the
density field which coincide with the parameter $\nu$ on account of
$\sigma$ being unity ($\rho_{\rm th} = \nu\sigma = \nu$).  $\nu$ is
used to label the levels and is related to the volume filling fraction
through the following equation:
%%%%%%%%%%%%%%%%%
\b FF_V(\nu) =
{1\over\sqrt{2\pi}}\int_{\nu}^{\infty}e^{-{t^2\over2}} dt.  
\e 
%%%%%%%%%%%%%%%%%
The samples are finally used for ensemble averaging.

The Minkowski Functionals (MFs) of a GRF are fully specified in terms
of a length-scale $\lambda_c$,
%%%%%%%%%%%%%%%%%
\b
\label{eq:lamb}
\lambda_c = \sqrt{{2\pi\xi(0)\over|\xi^{''}(0)|}};~~~~~\sigma^2 = \xi(0).
\e
%%%%%%%%%%%%%%%%%
$\lambda_c$ can be analytically derived from a knowledge of the power
spectrum. It can also be estimated numerically by evaluating the
variance $\xi$(0) of the field and the variance $\xi^{''}$(0) of any
of its first spatial derivatives (for more details see Matsubara 2003).

For a GRF in three dimensions the four MFs (per unit volume) are
\citep{tomita90, matsub03}:
%%%%%%%%%%%%%%%%%%
\b\label{eq:apMF1} 
V(\nu) = {1\over2} - {1\over2}\Phi({\nu\over\sqrt{2}});~~~\Phi(x) =
{2\over\sqrt{\pi}}\int_0^{x'} \exp\left(-{x'^2\over2}\right) dx',
\e
%%%%%%%%%%%%%%%%%%
where $\Phi$(x) is the error function.
\b\label{eq:apMF2} 
S(\nu) = {2\over\lambda_c}\sqrt{{2\over\pi}} \exp\left({-\nu^2\over2}\right), 
\e 
%%%%%%%%%%%%%%%%%%
\b\label{eq:apMF3}
C(\nu) = {\sqrt{2\pi}\over\lambda_c^2}\nu\exp\left({-\nu^2\over2}\right), 
\e
%%%%%%%%%%%%%%%%%%
\b \label{eq:apMF4}
G(\nu) =
{1\over\lambda_c^3\sqrt{2\pi}}(1-\nu^2)\exp\left({-\nu^2\over2}\right).
\e
%%%%%%%%%%%%%%%%%%

%%%%%%%%%%%%%%%%%%
\begin{figure}
\begin{center}
%  \centering
%  \centerline{
\end{center}
\resizebox{8cm}{!}{\includegraphics{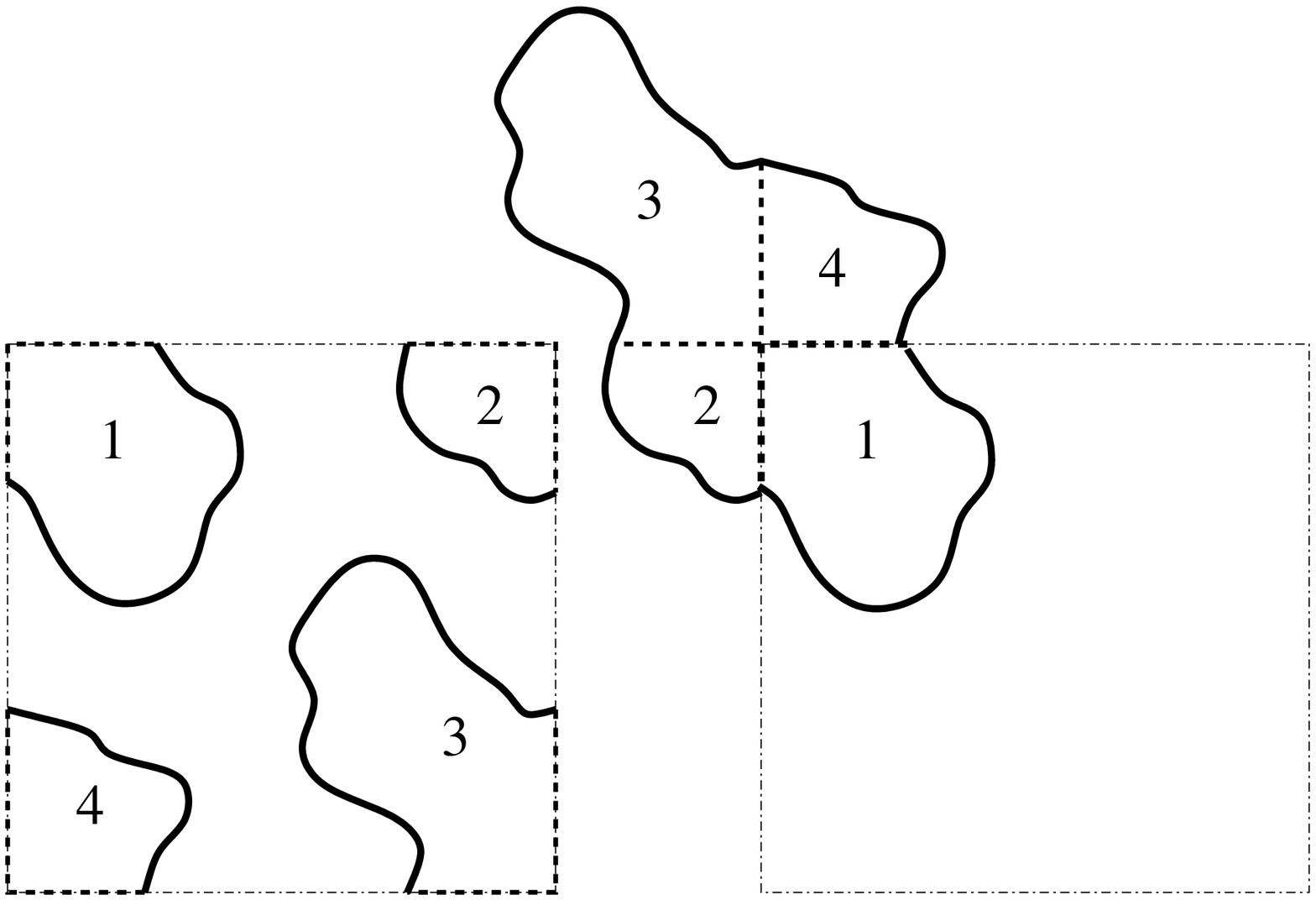}}
\caption{A generic 2$-$D Gaussian random field with four clusters touching
  the boundaries of the box is shown in the left panel. SURFGEN-like
  algorithm in 2$-$D would treat these as four different contours.
  However, due to the periodic nature of the GRF, these four clusters
  in fact constitute a single bigger cluster and the {\em real}
  boundaries of the cluster are as shown by bold lines in the right
  panel of the figure.  The dashed solid lines are interior to the
  cluster and are replaced by 2-dimensional planes in 3$-$D, which
  should not be considered in estimating the area and mean curvature.
  We take care of this effect while estimating MFs for GRFs.}
\label{fig:projection}
\end{figure}
%%%%%%%%%%%%%%%%%%
Let us now describe the boundary conditions which we adopt and which
incorporate the periodic nature of the GRFs generated using a FFT
routine.  GRFs defined on a grid are, by construction, periodic in
nature. As a result, two overdense excursion sets visibly separated
from each other inside the box and touching two opposite sides of the
box would constitute a single cluster. To make this more clear, let us
turn to Fig.~\ref{fig:projection} which shows
%which shows a generic
%case of a 2-D GRF at a sufficiently high threshold of density. There
four clusters (labelled 1, 2, 3 and 4 respectively) in the left panel.
The boundaries of these clusters $-$ as drawn by an analog of SURFGEN
in 2-D $-$ are shown in bold, and we note that the sides of the box
(bold dashed lines) are incorporated when we define the clusters.
However, due to the periodicity in the density field, these four
clusters are in fact one single cluster (right panel).  We note that
the portion of the contours due to the boundaries of the box occurs in
the interior of the actual cluster (shown with a bold dashed line).

It is easy to extrapolate this situation to 3$-$D. Note that the
interpolation method employed by SURFGEN will correctly estimate the
global volume of the excursion set even if we worked with the field
{\it as it is}. In 2$-$D this is same as estimating the area enclosed
by the {\em full} cluster without having to piece the four
``clusters'' (labelled by 1, 2, 3 and 4) together. Our estimation of
global genus will also turn out to be exact, and as it can be shown,
this will hold true even if either of these four constituent clusters
exhibits a nontrivial topology. On the other hand, if we work with the
same set of contours for estimating global area and mean curvature, we
will wrongly be adding an excess contribution due to the sides of the
box. (In 2$-$D, this translates to adding the contribution of the
dashed line segments to the perimeter of the contour in the right
panel of Figure \ref{fig:projection}).  The actual area and mean
curvature contribution comes from the interface between overdense and
underdense regions (such as the 2$-$D contours drawn in solid bold
line in both the panels) as opposed to the artificial boundaries of
the clusters (shown dashed).

We note that incorporating periodic boundary conditions in order to
construct a total contour as shown in the right panel of
Fig.~\ref{fig:projection} is nontrivial from the numerical point of
view. When the excursion set touches all the boundaries (as, for
example, at a very low thresold of density, below the percolation
threshold), the corresponding algorithm could further run into an
infinite loop, which makes it impractical for implementation. In the
light of this, it would be more manegeable if we restricted ourselves
to using contours which are closed at the boundaries of the box (like
the ones in the left panel of Fig.~\ref{fig:projection}), but devise a
way to exclude the {\em excess} contribution to surface area and mean
curvature from the boundaries of the box. A uniform prescription of
this type, when applied to all the appropriate clusters at all the
thresholds of density will lead to correct estimation of all the
global MFs. This then, is the choice of boundary conditions
incorporated in our calculation of a GRF. A futher cautionary note
concerns the familiar W-shape of the genus-curve. Recall that the
genus-curve peaks at $\nu=0$ signifying a sponge-like topology of the
medium at the mean density threshold.  Note further that
G($\nu=\pm\sqrt{3}$)= $-$2G(0)$\exp\left(-{3\over2}\right)$ are the
two negative minima of the genus curve.  In fact for $|\nu| >$ 1, the
genus is {\em always} negative, and approaches zero from below when
$|\nu| \gg 1$.  For $\nu< -1$, this negative genus refers to isolated
underdense regions (bubbles) in the overdense excursion set.
Similarly, the negative genus for $\nu > 1$ should be interpreted as
being caused by isolated overdense regions (meatballs) in the
underdense excursion set.
%Its clear from here that if we assess the topology of
%the overdense excursion set alone, its corresponding genus-curve would
%asymptotically vanish as $\nu\rightarrow\infty$. 
To summarize, in order to determine the genus for a GRF we need to 
estimate the genus of the {\em overdense}
excursion set for $\nu\le$0 and the genus 
of the {\em underdense} excursion set for $\nu > $ 0, thereby making use of the
symmetry property of the GRFs.
(Note however that in our earlier determination of
the genus curve for
N-body simulations in this paper, we have dealt with
the topology of overdense
regions only, because of which the genus curve is asymmetric 
with respect to $\nu = 0$ and is {\em always positive} when $\nu > 1$.)

Having elaborated on the nature of boundary conditions used
in our evaluation of MF's for a GRF, we now present our
results.

%%%%%%%%%%%%%%%%%
\begin{figure*}
  \centering \centerline{
    \includegraphics[width=4in]{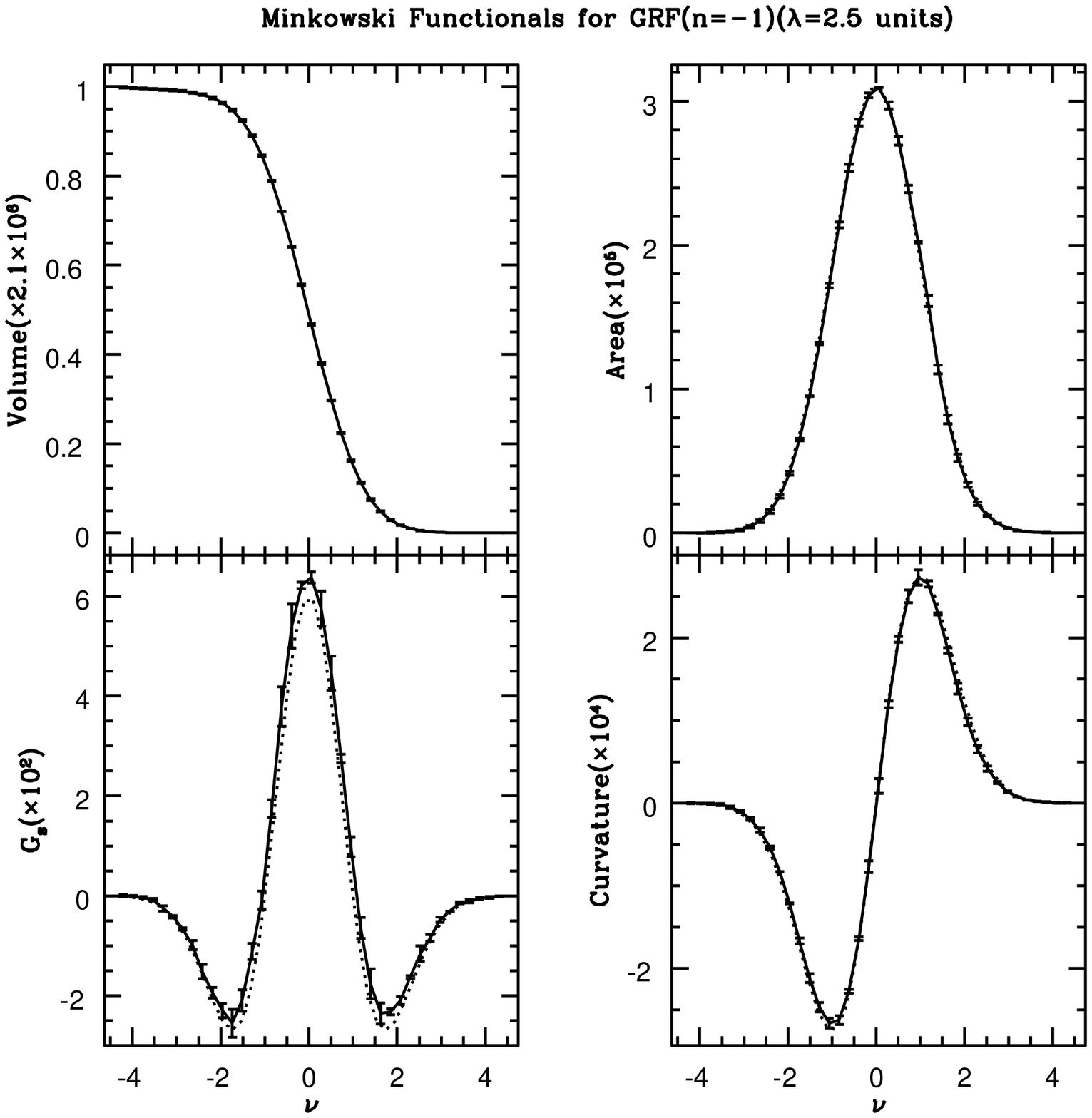} }
\caption{Minkowski Functionals for a GRF (P(k)$\sim$k$^{-1}$), smoothed
  with a gaussian filter with $\lambda$=2.5 grid-units. 
  Values of the four Minkowski fnctionals determined using SURFGEN are shown 
  as solid lines together with the 1$\sigma$
  scatter. Exact analytical results are shown as dotted lines.}
\label{fig:results2.5}
\end{figure*}
%%%%%%%%%%%%%%%%%

Figure \ref{fig:results2.5} shows the global MFs averaged over 3
realizations of the density field (solid lines with 1$\sigma$ error
bars) along with the exact analytical results (shown dotted). There
are in all 40 levels for every realization. 
Figure \ref{fig:results2.5} clearly shows the remarkable agreement between 
exact theoretical results (\ref{eq:apMF1}) - (\ref{eq:apMF4}) and 
numerical estimates obtained using SURFGEN.
We therefore conclude that SURFGEN determines the MF's of a GRF to great
precision.

The above discussion demonstrated the important role played by
boundary conditions 
in reproducing the analytical predictions for GRFs. Having tested the
performance of SURFGEN against GRFs, we are faced with the following choice 
of boundary conditions when dealing with N$-$body simulations and mock/real
galaxy catalogues: (i) we could either correct for the boundaries when
dealing with the overdense regions which encounter the faces of the
survey-volume,
%(an approach which is not advocated while dealing with
%mock or real catalogues due to the fact that the samples are not
%periodic by construction), 
or, (ii) we could avoid doing this and treat all
measurements with a tacit understanding that area and mean curvature contain an
excess contribution which arises because of boundary effects. 
Of the two options, the
former is computationally more demanding. In addition, its effect is
appreciable only significantly after the system has percolated, i.e., when
there are a large number of clusters touching the boundaries of the
box. The corresponding value of the filling factor (FF$_V$ $>$ FF$_{\rm
perc}$) is useful only when one wishes to compare two samples through the
trends in their global MFs. 
%This is because the structures that we
%find in this range are all fattened versions derived and extended out
%of the {\em real} structures; thus not ideally suitable for
%comparing various models. The {\em real} structural elements exist
%either at or above the percolation threshold, a situation where
%corrections due to boundaries are negligible. 
Since the structural elements of the cosmic web (superclusters, voids)
are usually identified near the percolation threshold, and are
therefore not very sensitive to boundary effects, we have chosen
option (ii) in our analysis of N-body simulations in this paper.
Option (ii) is also more suited for dealing with real galaxy
catalogues which do not satisfy periodic boundary conditions.  We
should also point out that since we compare two samples under the same
conditions $-$ the same volume and dimensions of the box, same
resolution, etc.$-$ the excess contribution to the area and mean
curvature caused by box boundaries enters in identical fashion for both
the samples.  We therefore prefer option (ii) to (i) and all our
simulations in Section 5 are analysed without correcting for
contributions from the boundaries.  (It may also be noted that the
contribution from the boundaries becomes less important as we deal
with larger survey volumes.)
\end{document}